\documentclass{article}
\usepackage[utf8]{inputenc}
\usepackage{hyperref}
\usepackage{physics, siunitx, graphicx, amsmath, amssymb, mathtools, amsfonts, pgfplots, caption, subcaption, epstopdf, booktabs}
\usepackage[version=4]{mhchem}
\usepackage{comment}
\usepackage{placeins}
\usepackage{tikz}
\usetikzlibrary{patterns,arrows}
\pgfplotsset{compat=1.13}

\usepackage{geometry}
\numberwithin{equation}{section}

\newcommand{\ts}[1]{_{\text{#1}}}
\newcommand{\kin}[1]{\text{k} \in \{\text{#1}\}}
\newcommand{\asy}[2]{#1^{(#2)}}
\newcommand{\hpp}{h''}
\newcommand{\hp}{h'}
\newcommand{\hppp}{h'''}
\newcommand{\sigmapp}{\sigma''}
\newcommand{\ve}{\varepsilon}
\newcommand{\av}[1]{\langle #1 \rangle}
\newcommand{\nablad}{\nabla_{\!\delta}}
\newcommand{\nablap}{\nabla_{\!\perp}}

\usepackage{amsfonts}
\usepackage{graphicx}
\usepackage{epstopdf}
\usepackage{algorithmic}
\ifpdf
  \DeclareGraphicsExtensions{.eps,.pdf,.png,.jpg}
\else
  \DeclareGraphicsExtensions{.eps}
\fi

\title{Asymptotic Reduction of a Lithium-ion Pouch Cell Model}

\author{Robert Timms${}^{*\text{a,b}}$
  $\quad$ Scott G.~Marquis${}^{\text{a}}$
  $\quad$ Valentin Sulzer${}^{\text{c}}$ \\
  $\quad$ Colin P.~Please${}^{\text{a,b}}$
  $\quad$ S.~Jonathan Chapman${}^{\text{a,b}}$ \\
   \footnotesize{${}^*$Corresponding author: \texttt{timms@maths.ox.ac.uk}} \\
  \footnotesize{${}^{\text{a}}$Mathematical Institute, Andrew Wiles Building, Woodstock Road, Oxford OX2 6GG, UK} \\
 \footnotesize{${}^{\text{b}}$The Faraday Institution, Quad One, Becquerel Avenue, Harwell Campus, Didcot, OX11 0RA, UK} \\
 \footnotesize{${}^{\text{c}}$Department of Mechanical Engineering, University of Michigan,}\\
  \footnotesize{2044 WE Lay Auto Lab, 1231 Beal Ave, Ann Arbor MI 48109-2133}
}

\date{}
\usepackage{amsopn}

\begin{document}

\maketitle

\begin{abstract}
  A three-dimensional model of a single-layer lithium-ion pouch cell is presented which couples conventional porous electrode theory describing cell electrochemical behaviour with an energy balance describing cell thermal behaviour. Asymptotic analysis of the model is carried out by exploiting the small aspect ratio typical of pouch cell designs. The analysis reveals the scaling that results in a distinguished limit, and highlights the role played by the electrical conductivities of the current collectors.
  The resulting model comprises a collection of one-dimensional models for the through-cell electrochemical behaviour which are coupled via two-dimensional problems for the Ohmic and thermal behaviour in the planar current collectors.  A further limit is identified which reduces the problem to a single volume-averaged through-cell model, greatly reducing the computational complexity. Numerical simulations are presented which illustrate and validate the asymptotic results.
\end{abstract}

\section{Introduction}
Lithium-ion batteries are one of the most widely used technologies for energy storage, with applications ranging from portable electronics to electric vehicles \cite{VanNoorden2014, scrosati2010, armand2008}. Due to their popularity, there is a continued interest in the development of mathematical models that can efficiently and accurately describe the behaviour observed during lithium-ion battery operation. Such models often provide a simplified one-dimensional description of the electrochemical behaviour in the through-cell direction, justified by the assumption that the behaviour in the remaining two dimensions is uniform.
However, larger-sized batteries, such as those used in the electric vehicle sector, exhibit non-uniform behaviour in the current and temperature distribution, which can adversely affect battery performance and lifetime \cite{kosch2018}. In particular, local variations in the temperature may lead to cells ageing in a non-uniform manner \cite{rieger2016}. A particularly striking example of non-uniform behaviour can be seen in Figure~2 of \cite{birkl2017degradation}, which depicts extremely non-uniform lithiation in the negative electrode of a pouch cell. There is a need to better understand the origins of this non-uniformity, that is, how material properties, cell geometry, and operating conditions can give rise to local changes in the cell potential and temperature.

Since the pioneering work of Newman \cite{newman1962, newman2012, doyle1993}, who developed a continuum description of porous electrode behaviour, there has been a large body of work devoted to the mathematical modelling of batteries \cite{gomadam2002,ramadesigan2012}. Not all existing models are one-dimensional---the multiscale and multidimensional nature of the problem has also been accounted for in the literature (e.g. \cite{kosch2018, farag2017, baker1999, hosseinzadeh2018, kim2008, northrop2015}), with many models based on extensions or adaptations of the porous electrode model developed by Doyle, Fuller and Newman \cite{doyle1993} (the DFN model), or reductions thereof.

While fully-coupled three-dimensional electrochemical and thermal models provide useful information for predicting cell behaviour, they are often too computationally expensive to be practically useful, and simplifications must be made. One approach is to treat the electrochemical problem as a network of resistors, coupled to a three-dimensional thermal model \cite{gerver2011, kim2011, lee2013}. A current-voltage relation is given for each of the resistors in the network, which can either be a complicated description based on porous electrode theory \cite{lee2013}, or a simplified description, such as a nonlinear resistor fitted to an electrochemical model \cite{gerver2011} or to data \cite{kim2008}. This approach  reduces the three-dimensional electrochemical model to system of one-dimensional electrochemical models   coupled via a two-dimensional electrical problem in the current collectors, and a three-dimensional thermal model across the entire cell.

Such simplifications, sometimes referred to as ``potential pair'' models, are usually made in an ad-hoc manner (e.g. \cite{kosch2018, gerver2011, kim2011, lee2013}). In this paper, we provide a systematic asymptotic reduction of a full  three-dimensional  pouch-cell model for large   current collector conductivity and small aspect ratio, identifying the parameter regimes in which such a reduction is possible.
We will find that there are two distinguished limits: one in which
the model reduces to a set of through-cell one-dimensional models coupled through a two-dimensional problem for the boundary conditions, and a second in which only a single through-cell one-dimensional problem needs to be solved, with an additional  two-dimensional problem needed to calculate an in-series resistance.
We focus our attention on a rectangular pouch-cell geometry, but a similar  analysis can be performed for other cell geometries, such as those found in cylindrical or prismatic cells.

The paper is laid out as follows. In Section~\ref{sec:model} we present the governing equations for the full three-dimensional model, expressed in terms of dimensionless variables. Full details of the dimensional model may be found in the supplementary material. In Section~\ref{sec:smallaspectratio}, we present an asymptotic reduction of the model, while in  Section~\ref{sec:results} we compare numerical solutions of the full and reduced models, and discuss the results. Finally, in Section~\ref{sec:conclusion}, we draw our conclusions.

\section{Model Equations}
\label{sec:model}
\begin{figure}
    \centering
    \includegraphics[width=\textwidth]{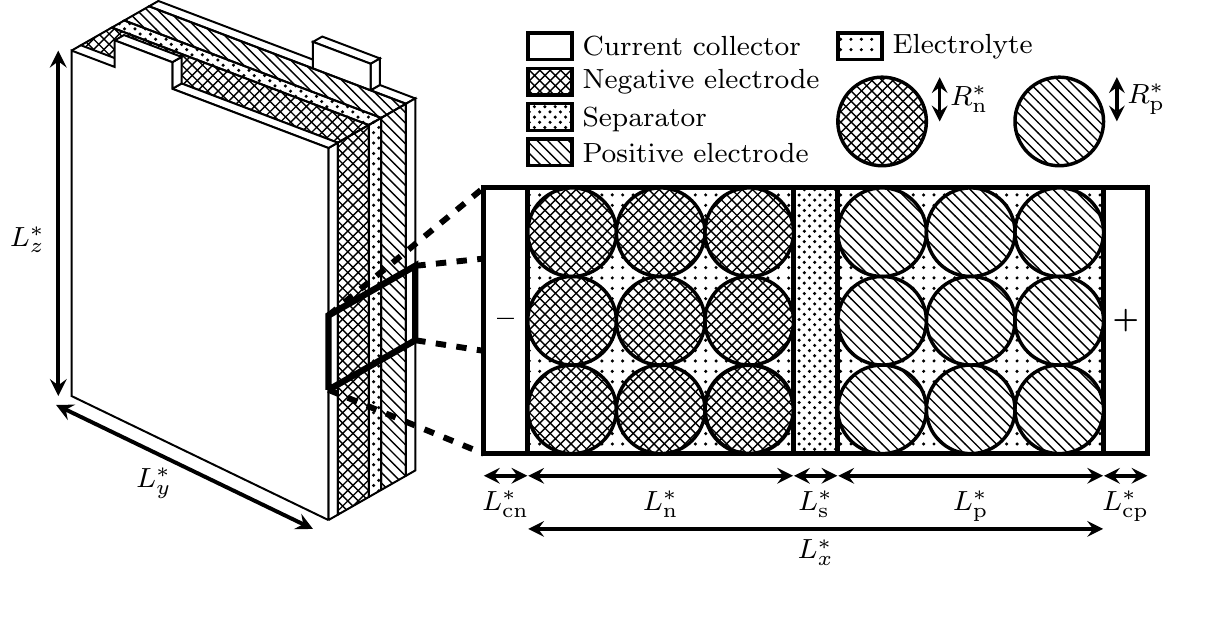}
\caption{Sketch of the three-dimensional pouch cell model.}
\label{fig:cell_sketch}
\end{figure}

We consider a single-layer lithium-ion pouch cell, which consists of both negative and positive current collectors, between which a negative electrode, a porous separator, and a positive electrode are sandwiched, as shown schematically in Figure~\ref{fig:cell_sketch}. Each electrode is a porous medium, comprising  active material particles, in which lithium is stored, held together with a binder. The binder material is electrically conducting and acts to maintain electrical connectivity between the active material particles and the current collectors. Both electrodes and the porous separator are flooded with electrolyte, which carries ionic charge. During operation, a current is drawn from the cell via tabs connected to each current collector, which are depicted as rectangular protrusions in Figure~\ref{fig:cell_sketch}. For readers unfamiliar with lithium-ion batteries and their construction, an excellent introduction can be found in \cite{plett2015battery}.

The electrochemical model developed here is based on the DFN model for porous electrodes \cite{newman1962, newman2012, doyle1993}. This model has become the standard choice for describing the behaviour of lithium-ion batteries \cite{fuller1994, newman2012, gomadam2002, ramadesigan2012}, and comprises equations for mass and charge conservation in the solid (active material and binder) and liquid (electrolyte) phases. Thermal effects may be incorporated by introducing equations describing the balance of thermal energy \cite{gu2000thermal, bernardi1985}. The model can be formally derived
from a microscopic description of electrochemical processes through  volume-averaging or homogenisation techniques \cite{wang1998, richardson2012, schmuck2015}, but we do not give the details here.

Often such homogenisation techniques result in a model posed on the macroscale with effective properties which depend on the microscale geometry. However, due to relatively slow diffusion in the particles, the DFN model retains a microscopic description of mass transport within the active material particles, which must be solved alongside macroscopic equations for charge transport in the solid material, and charge and mass transport within the electrolyte. In order to simplify the microscale model it is standard to assume that the active material particles may be treated as spheres, and that behaviour within the particles is spherically symmetric. We adopt this standard assumption, but note that the analysis could be easily extended to account for alternative particle shapes (e.g. \cite{Vanimisetti2012}). It is the combination of a one-dimensional macroscale equation coupled with a one-dimensional (radial) microscopic equation which gives rise to the alternative model name of ``pseudo-two-dimensional'' (often abbreviated to P2D).

It is straightforward to apply the principles of the DFN model to develop a full three-dimensional model of a lithium-ion pouch cell, as illustrated in Figure~\ref{fig:cell_sketch}. The model comprises the traditional DFN components of a porous negative electrode, separator, and positive electrode, each extended to three dimensions, as well as two additional components: a negative current collector and a positive current collector. We model the current collectors as Ohmic conductors. We assume a uniform current density is drawn from the positive tab, while the potential on the negative tab is  uniform and set to a reference value of \SI{0}{V}. The terminal voltage is then defined as the potential averaged over the positive tab. Because thermal effects are of particular interest in the study of pouch cells, we extend the DFN model to include an equation for energy conservation in each cell component, accounting for the various forms of heating and cooling which occur in the cell during operation \cite{bernardi1985}. The dimensionless model equations are summarised in Section~\ref{sec:model_eqns}, and are described in further detail in the supplementary material.

\subsection{Notation}
Before stating the governing equations we comment on our notation. Throughout we use a superscript $^*$ to denote dimensional quantities. We denote electric potentials by $\phi$, current densities by $\boldsymbol{i}$, lithium concentrations\footnote{In the electrolyte $c$ denotes the lithium-ion concentrations.} by $c$, molar fluxes by $\boldsymbol{N}$, and temperatures by $T$. To distinguish potential, fluxes and concentrations in the electrolyte from those in the solid phase of the electrode, we use a subscript $\text{e}$ for electrolyte variables and a subscript $\text{s}$ for solid phase variables. To indicate the region within which each variable is defined, we include an additional subscript k, which takes one of the following values: n (negative electrode), p (positive electrode), cn (negative current collector), cp (positive current collector), or s (separator).
For example, the notation $\phi^*\ts{s,n}$ refers to the dimensional electric potential in the solid phase of the negative electrode. When stating the governing equations, we take the region in which an equation holds to be implicitly defined by the subscript of the variables.
These regions, after nondimensionalisation, are given by
\begin{align*}
    \Omega\ts{cn} &=  [-L\ts{cn},0]\times\Omega, &
    \Omega\ts{n} &=  [0,L\ts{n}]\times\Omega, &
    \Omega\ts{s} &=  [L\ts{n},1-L\ts{p}]\times\Omega, \\
    \Omega\ts{p} &=  [1-L\ts{p},1]\times\Omega,&
    \Omega\ts{cp} &=  [1, 1+L\ts{cp}]\times\Omega,
    \end{align*}
corresponding to the negative current collector, negative electrode, separator, positive electrode, and positive current collector, respectively,
where   $\Omega = [0,L_y]\times[0,L_z]$
is the projection of the cell onto the $(y,z)$-plane.
For ease of reference we provide a glossary of the dimensionless variables, and their region of definition, in Table \ref{table:dimensionless_variables}.

The through-cell coordinate $x$ has been scaled with the distance between the current collectors (i.e.~with  $L_x^*=L\ts{n}^*+L\ts{s}^*+L\ts{p}^*$ shown in Fig.~\ref{fig:cell_sketch}), while the transverse coordinates $y$ and $z$ have been scaled with a typical transverse dimension $L^*$ (so that $L_y$ and $L_z$ are dimensionless constants of $O(1)$). We also introduce the notation $\partial\Omega\ts{tab,k}$ to refer to the negative and positive tabs ($\kin{cn, cp}$), $\partial\Omega\ts{ext,k}$ to refer to the external boundaries of region $\kin{cn, n, s, p, cp}$, and $\partial\Omega_{\text{k}_1, \text{k}_2}$ to refer to the interface between regions $\text{k}_1$ and $\text{k}_2$. For instance, the notation $\partial\Omega\ts{n,s}$ refers to the interface between the negative electrode and the separator. Finally, for $\kin{cn, cp}$  we use $\partial\Omega\ts{tab,k,$\perp$}$ to denote the projection of the tabs onto the $(y,z)$-plane, and
$\partial\Omega\ts{ext,k,$\perp$} =  \partial\Omega\setminus\partial\Omega\ts{tab,k,$\perp$}$ to denote the
non-tab region of the boundary of the projection.

\subsection{Governing equations}
\label{sec:model_eqns}
A full description of the dimensional model and its nondimensionalisation is given in the supplementary material.
Here, we summarise the dimensionless three-dimensional DFN model. In the following, we use the scaled gradient operator
\begin{equation}
    \nablad \equiv \pdv{}{x} \boldsymbol{e}_1 + \delta \pdv{}{y}  \boldsymbol{e}_2 + \delta \pdv{}{z} \boldsymbol{e}_3,
  \end{equation}
where $\boldsymbol{e}_i$ is the unit vector in the $i$\,th direction, and $\delta = L^*_x/L^*$ is the aspect ratio of the cell (which arises because of the different scaling in the $x$ and $y$, $z$ directions).
A number of nondimensional parameters appear in the equations. The definitions of these, along with typical values, are listed in Table \ref{table:dimensionless_parameter_values}.

The current in the electrodes and current collectors is given by Ohm's law, which along with charge conservation implies
\begin{subequations}\label{eqn:solid_charge_conservation}
    \begin{align}
           \nablad \cdot \boldsymbol{i}\ts{s,k} &= 0,  &&\kin{cn, cp}, \label{eqn:current-collector-conservation} \\
        \boldsymbol{i}\ts{s,k} &= - \sigma\ts{k} \nablad \phi\ts{s,k}, &&\kin{cn, cp}, \label{eqn:current-collector-ohms-law} \\
        \nablad \cdot \boldsymbol{i}\ts{s,k} &= -j\ts{k}, && \kin{n, p}, \label{eqn:solid-current-conservation} \\
        \boldsymbol{i}\ts{s,k} &= - \sigma\ts{k} \nablad \phi\ts{s,k}, && \kin{n, p}, \label{eqn:solid-ohms-law}
    \end{align}
    where the interfacial current density $j\ts{k}$ represents charge transfer between the active material and the electrolyte.
The boundary conditions are
    \begin{align}
       \phi\ts{s,cn} &= 0, \quad &&\boldsymbol{x} \in \partial \Omega\ts{tab,cn}, &&& \label{bc:negtab} \\
       \delta \int_{\partial  \Omega\ts{tab,cp}} \boldsymbol{i}\ts{s,cp}\cdot\boldsymbol{n} \dd{A} &= I\ts{app}, && &&& \label{bc:postab} \\
       \boldsymbol{i}\ts{s,k}\cdot\boldsymbol{n} &= 0, \quad\quad &&\boldsymbol{x} \in \partial \Omega\ts{ext,k} &&&\kin{cn, n, p, cp} \label{bc:solext} \\
       \boldsymbol{i}\ts{s,ck}\cdot\boldsymbol{n} &= \boldsymbol{i}\ts{s,k}\cdot\boldsymbol{n}, \quad &&\boldsymbol{x} \in \partial \Omega\ts{ck,k}, \quad &&&\kin{n, p}, \label{bc:solid-current-cc} \\
       \boldsymbol{i}\ts{s,k}\cdot\boldsymbol{n} &= 0, \quad &&\boldsymbol{x} \in \partial \Omega\ts{k,s}, \quad &&&\kin{n, p},  \label{bc:solid-current-separator}\\
       \phi\ts{s,ck} &= \phi\ts{s,k}, \quad &&\boldsymbol{x} \in \partial \Omega\ts{ck,k}, \quad &&& \kin{n, p}. \label{bc:solid-phi-cont}
    \end{align}
In addition, at the positive tab one we have the option of either
(i) assuming a uniform potential over the surface of the tab, the value of which is given by satisfying the integral constraint \eqref{bc:postab}, or (ii) assuming a  uniform current density over the surface of the tab.
In our later numerical examples we make the latter assumption, so that \eqref{bc:postab} can be replaced by
\begin{equation}
    \delta\boldsymbol{i}\ts{s,cp}\cdot\boldsymbol{n} = \frac{I\ts{app}}{A\ts{tab,cp}}, \quad \boldsymbol{x} \in \partial \Omega\ts{tab,cp},
\end{equation}
where $A\ts{tab,cp}$ is the surface area of the positive tab. The terminal voltage is then defined as the average of the potential over the positive tab region.
\end{subequations}

Mass conservation in the active material leads to a spherical diffusion problem in the particles where the boundary flux is given by the interfacial current density
\begin{subequations}\label{eqn:lithium_concentration_in_active_material}
    \begin{align}
      \mathcal{C}\ts{k}\pdv{c\ts{s,k}}{t} &= -\frac{1}{r\ts{k}^2}\pdv{r\ts{k}}\left(r\ts{k}^2N\ts{s,k}\right), \quad
      N\ts{s,k} = -D\ts{s,k}(c\ts{s,k}, T\ts{k})\pdv{c\ts{s,k}}{r\ts{k}} && \kin{n, p}, \label{eqn:electrode:molar:conservation} \\
        \label{eqn:electrode:molar:bc}
        N\ts{s,k}\big|_{r\ts{k}=0} &= 0, \qquad \qquad
        \frac{a\ts{k}\gamma\ts{k}}{\mathcal{C}\ts{k}}N\ts{s,k}\big|_{r\ts{k}=1} = j\ts{k}, && \kin{n, p}, \\
        c\ts{s,k}\big|_{t=0} &= c\ts{s,k,$0$},  && \kin{n, p},
    \end{align}
\end{subequations}
where $r\ts{k}$ is the microscale coordinate indicating radial position in the particle.

Similarly, we have charge conservation in the electrolyte in which the current is described using a modified Ohm's law:
\begin{subequations}\label{eqn:current_in_electrolyte}
    \begin{align}
        \nablad \cdot \boldsymbol{i}\ts{e,k} &=
        \begin{cases}
            j\ts{k}, \quad &\text{k}=\text{n, p}, \\
            0, \quad & \text{k} = \text{s},
        \end{cases} &&\kin{n, s, p}, \label{eqn:electrolyte:current-conservation} \\
        \mathcal{C}\ts{e} \boldsymbol{i}\ts{e,k} &= \epsilon\ts{k}^{\text{b}} \hat{\kappa}\ts{e} \kappa\ts{e}(c\ts{e,k}, T\ts{k}) \left(   2(1-t^+)(1 + \Theta T\ts{k}) \nablad\left(\log c\ts{e,k}\right) - \nablad\phi\ts{e,k}\right), &&\kin{n, s, p}.\label{eqn:McInnesEqn}
    \end{align}
The current in the electrolyte satisfies the no flux boundary conditions
    \begin{align}
        \label{eqn:electroltye:bc1}
        \boldsymbol{i}\ts{e,k}\cdot\boldsymbol{n} &= 0, \quad \boldsymbol{x} \in \partial \Omega\ts{ck,k}, \quad &&\kin{n, p}, \\
        \label{eqn:electroltye:bc2}
        \boldsymbol{i}\ts{e,k}\cdot\boldsymbol{n} &= 0,  \quad \boldsymbol{x} \in \partial \Omega\ts{ext,k},  &&\kin{n, s, p}.
    \end{align}
\end{subequations}
Mass conservation in the electrolyte leads to a reaction-diffusion equation for the lithium-ion concentration:
\begin{subequations}\label{eqn:lithium_ion_concentration_in_electrolyte}
    \begin{align}
        \mathcal{C}\ts{e} \gamma\ts{e} \epsilon\ts{k} \pdv{c\ts{e,k}}{t} &= -\gamma\ts{e}\nablad \cdot \boldsymbol{N}\ts{e,k} +  \mathcal{C}\ts{e} \nablad\cdot\boldsymbol{i}\ts{e,k}, &&\kin{n, p}, \\ \label{eqn:electrolyte:molar:conservation}
    \boldsymbol{N}\ts{e,k} &= -\epsilon\ts{k}^{\text{b}}  D\ts{e}(c\ts{e,k}, T\ts{k}) \nablad c\ts{e,k} + \frac{\mathcal{C}\ts{e} t^+}{\gamma\ts{e}} \boldsymbol{i}\ts{e,k}, &&\kin{n, p},
    \end{align}
which must satisfy the initial and boundary conditions
    \begin{align}
    \label{eqn:electrolyte:molar:bc2}
    \boldsymbol{N}\ts{e,k}\cdot\boldsymbol{n} &= 0, \quad \boldsymbol{x} \in \partial \Omega\ts{ext,k}, \quad && \kin{n, s, p}, \\
    \label{eqn:electrolyte:molar:bc1}
    \boldsymbol{N}\ts{e,k}\cdot\boldsymbol{n} &= 0, \quad \boldsymbol{x} \in \partial \Omega\ts{ck,k}, \quad && \kin{n, p}, \\
     \label{eqn:electrolyte:molar:bc4}
    \boldsymbol{N}\ts{e,k}\cdot\boldsymbol{n} &= \boldsymbol{N}\ts{e,s}\cdot\boldsymbol{n}, \quad \boldsymbol{x} \in \partial \Omega\ts{k,s}, \quad && \kin{n, p}, \\
    \label{eqn:electrolyte:molar:bc3}
     c\ts{e,k} &= c\ts{e,s} \quad \boldsymbol{x} \in \partial \Omega\ts{k,s}, \quad && \kin{n, p}, \\
    c\ts{e,k}\big|_{t=0} &= c\ts{e,$0$}, \quad && \kin{n, s, p}.
    \end{align}
\end{subequations}
The electrochemical reactions at the surface of the solid particles are described by symmetric Butler-Volmer kinetics \cite{newman2012}:
\begin{subequations}\label{eqn:electrochemical_reactions}
    \begin{align}
        j\ts{k} &=  j\ts{$0,$k} \sinh\left(\frac{\eta\ts{k}}{2(1 + \Theta T\ts{k})} \right), &&\kin{n, p}, \\
     j\ts{$0,$k} &= \frac{\gamma\ts{k}}{\mathcal{C}\ts{r,k}} m\ts{k}(T\ts{k}) c\ts{s,k}^{1/2} (1-c\ts{s,k})^{1/2}c\ts{e,k}^{1/2}\bigg|_{r\ts{k}=1}, && \kin{n, p}, \\
     \eta\ts{k} &= \phi\ts{s,k} - \phi\ts{e,k} - U\ts{k}(c\ts{s,k},c\ts{e,k},T\ts{k})\big|_{r\ts{k}=1}, && \kin{n, p}.
    \end{align}
\end{subequations}
Finally, we have an equation for energy conservation. This accounts for Ohmic heating in both the solid and electrolyte, as well as reversible and irreversible heating due to electrochemical reactions:
\begin{subequations}
  \label{thermaleqns}
        \begin{align}
        \delta^2 \mathcal{C}\ts{th}\rho\ts{k} \pdv{T\ts{k}}{t} = \nablad\cdot&\left(\lambda\ts{k} \nablad T\ts{k}\right) + \delta^2\mathcal{B}\left( Q\ts{ohm,k} + Q\ts{rxn,k} + Q\ts{rev,k} \right), \! &&\kin{cn, n, s, p, cp}, \label{eqn:thermal:cell}\\
     \label{eqn:thermal:ohm}
     Q\ts{Ohm,k} &= - \left(\boldsymbol{i}\ts{s,k} \cdot\nablad\phi\ts{s,k} + \boldsymbol{i}\ts{e,k} \cdot\nablad \phi\ts{e,k}\right),&&\kin{n, p}, \\
          \label{eqn:thermal:s}
          Q\ts{Ohm,s} &= -\boldsymbol{i}\ts{e,s}\cdot\nablad\phi\ts{e,s}, && \\
          \label{eqn:thermal:cc}
        Q\ts{Ohm,k} &= -\boldsymbol{i}\ts{s,k}\cdot\nablad\phi\ts{s,k}, &&\kin{cn, cp}    \\
     \label{eqn:thermal:rxn}
    Q\ts{rxn,k} &=  j\ts{k} \eta\ts{k}, &&\kin{n,p}, \\
    \label{eqn:thermal:rev}
    Q\ts{rev,k} &=  j\ts{k}(\Theta^{-1} + T\ts{k}) \pdv{U\ts{k}}{T\ts{k}}\bigg|_{T\ts{k}=0}, &&\kin{n, p},
        \end{align}
with $Q\ts{rxn,k}=Q\ts{rev,k}=0$ for $\kin{cn, s, cp}$.
We assume Newton cooling on all boundaries (including the tabs), but we allow for the cooling coefficient $h$ to vary spatially. Prescribing a uniform initial temperature the boundary and initial conditions are then
    \begin{align}
    \label{eqn:thermal:newtoncooling}
    -\lambda\ts{k}\nablad T\ts{k} \cdot \boldsymbol{n} &= h(\boldsymbol{x})T\ts{k}, \quad \boldsymbol{x} \in \partial \Omega\ts{ext}, &&\kin{cn, n, s, p, cp}, \\
    T\ts{ck} = T\ts{k}, \quad \lambda\ts{ck}\nablad T\ts{ck} \cdot\boldsymbol{n} &= \lambda\ts{k}\nablad T\ts{k}\cdot\boldsymbol{n}, \quad \boldsymbol{x} \in \partial \Omega\ts{ck,k}, \quad &&\kin{n, p}, \\
    T\ts{k} = T\ts{s}, \quad \lambda\ts{k}\nablad T\ts{k} \cdot\boldsymbol{n} &= \lambda\ts{s}\nablad T\ts{s}\cdot\boldsymbol{n}, \quad \boldsymbol{x} \in \partial \Omega\ts{k,s}, \quad &&\kin{n, p}, \\
    T\ts{k} \big|_{t=0} &= T\ts{0}, &&\kin{cn, n, s, p, cp}.
    \end{align}
    Some authors have considered the effects of different cooling scenarios, such as tab cooling vs. surface cooling, on battery operation. For instance, Hunt \emph{et al.} \cite{hunt2016} conducted experiments showing that surface cooling can lead to a greater loss of capacity compared with tab cooling when discharging cells at high rates. These different cooling scenarios can be investigate by choosing particular cooling functions $h$ (for example, $h$ may be larger on the tabs).

\end{subequations}

\section{Asymptotic analysis}\label{sec:smallaspectratio}
In this section, we exploit the small aspect ratio of a typical pouch cell by considering the limit $\delta \to 0$. This is similar to the approach taken in \cite{sulzer_thesis}. To enable a balance of terms in the current conservation equations we  rescale the transverse currents by writing
\begin{equation}
  \boldsymbol{i}\ts{s,k} = i\ts{s,k,1}\boldsymbol{e}_1 + \frac{\boldsymbol{i}\ts{s,k,$\perp$}}{\delta}, \quad \kin{cn, n, p, cp}, \quad \boldsymbol{i}\ts{e,k} = i\ts{e,k,1}\boldsymbol{e}_1 + \frac{\boldsymbol{i}\ts{e,k,$\perp$}}{\delta}, \quad \kin{n, s, p},
  \label{rescalecurrent}
\end{equation}
where $\boldsymbol{i}\ts{s,k,$\perp$}$ and $\boldsymbol{i}\ts{e,k,$\perp$}$ are the $y$-$z$ components of the solid and electrolyte current, respectively. For notational convenience, we define
\begin{equation}
  \nablap \equiv \pdv{y}\boldsymbol{e}_2 + \pdv{z}\boldsymbol{e}_3.
  \label{transversegrad}
\end{equation}

\subsection{The large conductivity limit}\label{sec:high_cond}
We consider the physically-relevant limit of large (dimensionless) conductivity in the current collectors. There is a distinguished limit when  $\sigma\ts{k} = \sigma\ts{k}'/\delta^2$ for $\kin{cn, cp}$, where $\sigma\ts{k}' = \mathcal{O}(1)$ as $\delta \rightarrow 0$. Further, to retain both heat loss from the current collector surfaces (area $O(1)$) and heat loss from the cell edges including the tabs (area $O(\delta)$) at leading order
we consider the limit in which $h = \delta^2 \hp$ for $\boldsymbol{x} \in \{-L\ts{cn},1+L\ts{cp}\}\times \Omega$ , and
$h = \delta \hpp$ for the remaining external boundaries, with $\hp$ and $\hpp$ of $\mathcal{O}(1)$ as $\delta \rightarrow 0$. Such a scaling for the heat transfer coefficient is applicable for cooling under free convection, but $h$ may be considerably larger for forced cooling (e.g. \cite{incropera07,carslaw59,bizeray16}).
All other parameters are taken to be $\mathcal{O}(1)$.

We expand each variable in powers of $\delta^2$ as $\delta \rightarrow 0$  in the form
\begin{equation}
\phi\ts{s,k} = \asy{\phi\ts{s,k}}{0} + \delta^2 \asy{\phi\ts{s,k}}{2} + \cdots.
\end{equation}

\subsubsection{Charge conservation in the solids}
Using (\ref{rescalecurrent})-(\ref{transversegrad}) the governing equations for charge transport in the current collectors, \eqref{eqn:current-collector-conservation}-\eqref{eqn:current-collector-ohms-law}, read
\begin{subequations}
\begin{align}
\label{eq:ccsolcurrent}
\pdv{i\ts{s,k,1}}{x} + \nablap \cdot \boldsymbol{i}\ts{s,k,$\perp$} &= 0, \quad &&\kin{cn, cp}, \\
\label{eq:ohmssolidck}
\delta^2 i\ts{s,k,1} &= -\sigma\ts{k}'\pdv{\phi\ts{s,k}}{x}, \quad \boldsymbol{i}\ts{s,k,$\perp$} = -\sigma\ts{k}'\nablap \phi\ts{s,k}, \quad &&\kin{cn, cp},
\end{align}
\end{subequations}
along with the rescaled boundary conditions at the tabs
\begin{subequations}
\begin{align}
\label{bc:negtabscaled}
\phi\ts{s,cn} &= 0, \quad &&\boldsymbol{x} \in \partial \Omega\ts{tab,cn}, &&& \\
\label{bc:postabscaled}
\boldsymbol{i}\ts{s,cp,$\perp$}\cdot\boldsymbol{n} &= \frac{I\ts{app}}{A\ts{tab,cp}}, \quad &&\boldsymbol{x} \in \partial \Omega\ts{tab,cp}, &&&
\end{align}
\end{subequations}
the no flux conditions \eqref{bc:solext}--\eqref{bc:solid-current-separator}, and continuity of the potential and current at the electrode/separator interfaces $x = L\ts{n}, 1 - L\ts{p}$. After expanding in powers of $\delta^2$, we immediately see from \eqref{eq:ohmssolidck} that $\asy{\phi\ts{s,cn}}{0}$ and $\asy{\phi\ts{s,cp}}{0}$ are independent of $x$, and therefore $\asy{\boldsymbol{i}\ts{s,cp,$\perp$}}{0}$ and $\asy{\boldsymbol{i}\ts{s,cp,$\perp$}}{0}$
are also independent of $x$. Then, integration of the leading-order terms in \eqref{eq:ccsolcurrent} and application of the appropriate boundary conditions gives
\begin{equation}
\label{eqn:cc:current}
L\ts{cn}\nablap\cdot \asy{\boldsymbol{i}\ts{s,cn,$\perp$}}{0} = -\mathcal{I}\ts{n}(y,z),  \qquad  L\ts{cp}\nablap \cdot \asy{\boldsymbol{i}\ts{s,cp,$\perp$}}{0} = \mathcal{I}\ts{p}(y,z)\qquad (y,z) \in \Omega,
\end{equation}
where $\mathcal{I}\ts{n}$ and $\mathcal{I}\ts{p}$ are the leading-order currents densities through the electrode/current collector interfaces:
\begin{equation}
\label{eqn:I_def}
\mathcal{I}\ts{n}(y,z) := \asy{i\ts{s,cn,1}}{0}\big|_{x=0} = \asy{i\ts{s,n,1}}{0}\big|_{x=0}, \qquad \mathcal{I}\ts{p}(y,z) := \asy{i\ts{s,cp,1}}{0}\big|_{x=1} = \asy{i\ts{s,p,1}}{0}\big|_{x=1}.
\end{equation}
Using (\ref{eq:ohmssolidck}) to write  (\ref{eqn:cc:current}) in terms of  potentials, gives
\begin{subequations}
\label{DFNCC:curr-coll}
\begin{equation}
    \label{DFNCC:eqn:phi_s_cncp}
    L\ts{cn}\sigma\ts{cn}'\nablap^2 \asy{\phi\ts{s,cn}}{0} = \mathcal{I}\ts{n},
    \qquad
    L\ts{cp}\sigma\ts{cp}'\nablap^2 \asy{\phi\ts{s,cp}}{0} = -\mathcal{I}\ts{p} \qquad (y,z) \in \Omega
\end{equation}
with boundary conditions
\begin{align}
\asy{\phi\ts{s,cn}}{0} &= 0 &\quad &(y,z)\in \partial \Omega\ts{tab,cn,$\perp$}, \label{bc:nDim:reference_potential} \\
-\sigma\ts{cp}' \nablap\asy{\phi\ts{s,cp}}{0}\cdot\boldsymbol{n} &= \frac{I_{\text{app}}}{A\ts{tab,cp}}& \quad &(y,z) \in \partial \Omega\ts{tab,cp,$\perp$}, \label{bc:nDim:applied current}\\
\nablap\asy{\phi\ts{s,k}}{0}\cdot\boldsymbol{n} &= 0 &\quad &(y,z) \in \partial \Omega\ts{ext,k,$\perp$}, \quad \kin{cn, cp}.
\end{align}
\end{subequations}
In the electrodes, after using (\ref{rescalecurrent})-(\ref{transversegrad}) the  governing equations \eqref{eqn:solid-current-conservation}-\eqref{eqn:solid-ohms-law} read
\begin{subequations}
\begin{align}
\label{eq:solidcurrent}
\pdv{i\ts{s,k,1}}{x} + \nablap \cdot \boldsymbol{i}\ts{s,k,$\perp$} &= -j\ts{k}, \quad &&\kin{n, p}, \\
\label{eq:ohmssolidk}
i\ts{s,k,1} &= -\sigma\ts{k}\pdv{\phi\ts{s,k}}{x}, \quad \boldsymbol{i}\ts{s,k,$\perp$} = -\delta^2\sigma\ts{k}\nablap \phi\ts{s,k}, \quad &&\kin{n, p},
\end{align}
\end{subequations}
At leading order $\asy{\boldsymbol{i}\ts{s,k,$\perp$}}{0} = 0$ for $\kin{n, p}$, so that
\begin{subequations}
\label{eqn:solid:current:np:full-problem}
    \begin{gather}
        \label{eqn:solid:current:np}
        \pdv{\asy{i\ts{s,k,1}}{0}}{x} = -\asy{j\ts{k}}{0}, \quad
        \asy{i\ts{s,k,1}}{0} = - \sigma\ts{k} \pdv{\asy{\phi\ts{s,k}}{0}}{x}, \quad \kin{n, p},
    \end{gather}
    with the boundary conditions
    \begin{align}
        \asy{\phi\ts{s,n}}{0}\big|_{x=0} &= \asy{\phi\ts{s,cn}}{0}, & \asy{i\ts{s,n,1}}{0}\big|_{x=L\ts{n}}& = 0, \\
        \asy{\phi\ts{s,p}}{0}\big|_{x=1}& = \asy{\phi\ts{s,cp}}{0}, &  \asy{i\ts{s,p,1}}{0}\big|_{x=1-L\ts{p}}&= 0.
    \end{align}
\end{subequations}

\subsubsection{Charge conservation in the electrolyte}
A similar calculation holds for charge conservation in the electrolyte. Using (\ref{rescalecurrent})-(\ref{transversegrad}) in \eqref{eqn:McInnesEqn} the transverse current in the electrolyte is
  \begin{align*}
     \boldsymbol{i}\ts{e,k,$\perp$} &= \delta^2 \epsilon\ts{k}^{\text{b}} \hat{\kappa}\ts{e} \kappa\ts{e}(c\ts{e,k}, T\ts{k}) \left( -  \nablap\phi\ts{e,k} + 2(1-t^+)(1 + \Theta T\ts{k}) \nablap\left(\log c\ts{e,k}\right)\right), \\
   \nonumber    &\hspace{8cm} \kin{n, s, p}, 
\end{align*}
Thus, to leading order in $\delta$, we have
$\asy{\boldsymbol{i}\ts{e,k,$\perp$}}{0} = 0$ for $\kin{n, s, p}$, and the flow of current in the electrolyte is also predominantly in the $x$-direction.
Then, at leading order in $\delta$,  equations (\ref{eqn:current_in_electrolyte}) give
\begin{subequations}
\label{DFNCC:eleclyte}
\begin{gather}
\label{eqn:reduced:electrolyte:current}
    \pdv{\asy{i\ts{e,k,1}}{0}}{x} = \begin{cases}
                        \asy{j\ts{k}}{0}, \quad \text{k} = \text{n, p},\\
                        0, \qquad \text{k} = \text{s},
                        \end{cases} \quad \kin{n, s, p}, \\
   \asy{i\ts{e,k,1}}{0} = \epsilon\ts{k}^{\text{b}} \hat{\kappa}\ts{e} \kappa\ts{e}(\asy{c\ts{e,k}}{0}, \asy{T\ts{k}}{0}) \left( -  \pdv{\asy{\phi\ts{e,k}}{0}}{x} + 2(1-t^+)(1 + \Theta \asy{T\ts{k}}{0}) \pdv{x}\left(\log \asy{c\ts{e,k}}{0}\right)\right),\\
\nonumber  \hspace{8cm}
    \quad  \kin{n, s, p},
\end{gather}
with boundary conditions given by \eqref{eqn:electroltye:bc1}--\eqref{eqn:electroltye:bc2}, and continuity conditions given by
\begin{align}
\label{bc:nDim:no_electrolyte_cc_current}
\asy{i\ts{e,n,1}}{0}\big|_{x=0} &= 0, &\quad \asy{i\ts{e,p,1}}{0}\big|_{x=1}&=0,  \\
\label{bc:nDim:charge_continuity_Ln}
\asy{\phi\ts{e,n}}{0}\big|_{x=L\ts{n}} &= \asy{\phi\ts{e,s}}{0}\big|_{x=L\ts{n}}, &\quad \asy{i\ts{e,n,1}}{0}\big|_{x=L\ts{n}} &= \asy{i\ts{e,s,1}}{0}\big\vert_{x=L\ts{n}}, \\
\label{bc:nDim:charge_continuity_Lp}
\asy{\phi\ts{e,s}}{0}\big|_{x=1-L\ts{p}}& = \asy{\phi\ts{e,p}}{0}\big|_{x=1-L\ts{p}}, &\quad
\asy{i\ts{e,s,1}}{0}\big|_{x=1-L\ts{p}} &= \asy{i\ts{e,p,1}}{0}\big|_{x=1-L\ts{p}} .
\end{align}
Note that (\ref{eqn:solid:current:np}), (\ref{eqn:reduced:electrolyte:current}) imply $\asy{i\ts{s,k,1}}{0}+ \asy{i\ts{e,k,1}}{0}$ is independent of $x$, so that
  \begin{equation}
    \asy{i\ts{s,k,1}}{0}+\asy{i\ts{e,k,1}}{0} =  \mathcal{I}\ts{k} \qquad  \kin{n,  p},
    \label{eq:conserve-charge}
\end{equation}
\end{subequations}
which can be used to eliminate $\asy{i\ts{s,k,1}}{0}$ in \eqref{eqn:solid:current:np}.
Note also that integrating \eqref{eqn:reduced:electrolyte:current} in $x$ and using  \eqref{bc:nDim:no_electrolyte_cc_current}-\eqref{eq:conserve-charge} gives
\begin{equation*}
 \mathcal{I}\ts{n} = L\ts{n}\asy{\bar{j}\ts{n}}{0} = \asy{i\ts{e,s,1}}{0} =  -L\ts{p}\asy{\bar{j}\ts{p}}{0} =\mathcal{I}\ts{p}= \mathcal{I},
\end{equation*}
say, where $\mathcal{I}=\mathcal{I}(y,z)$  is the through-cell current density,
and
\begin{equation*}
  \asy{\bar{j}}{0}\ts{n} = \frac{1}{L\ts{n}}\int_0^{L\ts{n}} \asy{j\ts{n}}{0} \dd{x}, \qquad \asy{\bar{j}\ts{p}} {0}= \frac{1}{L\ts{p}}\int_{1-L\ts{p}}^{L\ts{p}} \asy{j\ts{p}}{0} \dd{x},
\end{equation*}
are the electrode $x$-averaged leading-order interfacial current densities.

\subsubsection{Lithium conservation}
For the lithium concentrations in the solid and electrolyte we find at leading order
\begin{subequations}
\label{DFNCC:molar-cons}
\begin{align}
  \mathcal{C}\ts{k}\pdv{\asy{c\ts{s,k}}{0}}{t} &= -\frac{1}{r\ts{k}^2}\pdv{r\ts{k}}\left(r\ts{k}^2 \asy{N\ts{s,k}}{0}\right), \quad && \kin{n, p},\\
   \asy{N\ts{s,k}}{0} &= -D\ts{s,k}(\asy{c\ts{s,k}}{0}, \asy{T\ts{k}}{0}) \pdv{\asy{c\ts{s,k}}{0}}{r\ts{k}}, \quad && \kin{n, p}, \\
    \mathcal{C}\ts{e} \epsilon\ts{k} \gamma\ts{e} \pdv{\asy{c\ts{e,k}}{0}}{t} &= -\gamma\ts{e}\pdv{\asy{N\ts{e,k}}{0}}{x} + \mathcal{C}\ts{e} \pdv{\asy{i\ts{e,k}}{0}}{x}, \quad &&\kin{n, s, p}, \\
    \asy{N\ts{e,k}}{0} &= -\epsilon\ts{k}^{\text{b}}  D\ts{e}(\asy{c\ts{e,k}}{0}, \asy{T\ts{k}}{0})\pdv{\asy{c\ts{e,k}}{0}}{x} + \frac{\mathcal{C}\ts{e} t^+}{\gamma\ts{e}} \asy{i\ts{e,k}}{0} , && \kin{n, s, p},
\end{align}
with boundary conditions
\begin{align}
\label{bc:nDim:particle}
&\asy{N\ts{s,k}}{0}\big|_{r\ts{k}=0} = 0,& \quad \frac{a\ts{k}\gamma\ts{k}}{\mathcal{C}\ts{k}}\asy{N\ts{s,k}}{0}\big|_{r\ts{k}=1} &= \asy{j\ts{k}}{0}, \quad \kin{n, p}, \\
\label{bc:nDim:no_electrolyte_cc_flux}
&\asy{N\ts{e,n}}{0}\big|_{x=0} = 0, &\quad \asy{N\ts{e,p}}{0}\big|_{x=1}&=0,  \\
\label{bc:nDim:concentration_continuity_Ln}
&\asy{c\ts{e,n}}{0}\big|_{x=L\ts{n}} = \asy{c\ts{e,s}}{0}|_{x=L\ts{n}},& \quad \asy{N\ts{e,n}}{0}\big|_{x=L\ts{n}}&=\asy{N\ts{e,s}}{0}\big|_{x=L\ts{n}}, \\
\label{bc:nDim:concentration_continuity_Lp}
&\asy{c\ts{e,s}}{0}|_{x=1-L\ts{p}}=\asy{c\ts{e,p}}{0}|_{x=1-L\ts{p}},& \quad \asy{N\ts{e,s}}{0}\big|_{x=1-L\ts{p}}&=\asy{N\ts{e,p}}{0}\big|_{x=1-L\ts{p}},
\end{align}
and  initial conditions
    \begin{align}
    \label{ic:nDim:solid}
	\asy{c\ts{s,k}}{0}(x,y,z,r,0) &= c\ts{s,k,0}, && \kin{n, p},\\
	\label{ic:nDim:electrolyte}
	\asy{c\ts{e,k}}{0}(x,y,z,0) &= 1, && \kin{n, s, p}.
\end{align}
\end{subequations}

\subsubsection{Electrochemistry}
At leading order in $\delta$, the electrochemical reactions are given by
\begin{subequations}
\label{DFNCC:elecchem}
\begin{align}
\asy{j\ts{k}}{0} &=  \asy{j\ts{$0,$k}}{0} \sinh\left(\frac{ \asy{\eta\ts{k}}{0}}{2(1 + \Theta \asy{T\ts{k}}{0})} \right), && \kin{n, p}, \\
\asy{j\ts{$0,$k}}{0} &= \frac{\gamma\ts{k}}{\mathcal{C}\ts{r,k}}m\ts{k}(\asy{T\ts{k}}{0})(\asy{c\ts{s,k}}{0})^{1/2} (1 - \asy{c\ts{s,k}}{0})^{1/2}(\asy{c\ts{e,k}}{0})^{1/2}\bigg|_{r\ts{k}=1} && \kin{n, p}, \\
\asy{\eta\ts{k}}{0} &= \asy{\phi\ts{s,k}}{0} - \asy{\phi\ts{e,k}}{0} - U\ts{k}(\asy{c\ts{s,k}}{0},\asy{c\ts{e,k}}{0},\asy{T\ts{k}}{0})\big|_{r\ts{k}=1}, && \kin{n, p}.
\end{align}
\end{subequations}

\subsubsection{Energy conservation}
At leading order in (\ref{thermaleqns}) we find
\begin{equation}
\pdv[2]{\asy{T\ts{k}}{0}}{x} = 0, \quad \kin{cn, n, s, p, cp}
\end{equation}
with $\asy{T\ts{k}}{0}$ and $\lambda\ts{k} {\partial \asy{T\ts{k}}{0}}/{\partial x}$ continuous at $x=0$, 1, $L\ts{n}$ and $1-L\ts{p}$, and
    \begin{align}
      \pdv{\asy{T\ts{cn}}{0}}{x} \bigg|_{x=-L\ts{cn}} = \pdv{\asy{T\ts{cp}}{0}}{x} \bigg|_{x=1+L\ts{cp}} = 0,
    \end{align}
    giving $\asy{T}{0} = \asy{T}{0}(y,z,t)$, where we can drop the subscript k on the leading-order temperature since it is $x$-independent and the same across
     all of the cell components. At the next order, we find
     \begin{align}
  \mathcal{C}\ts{th} \rho\ts{k} \pdv{\asy{T}{0}}{t}& = \lambda\ts{k} \left( \pdv[2]{\asy{T\ts{k}}{2}}{x} + \nablap^2 \asy{T}{0} \right) +  \mathcal{B}\left( \asy{Q\ts{Ohm,k}}{0} + \asy{Q\ts{rxn,k}}{0} + \asy{Q\ts{rev,k}}{0} \right),\\
  \nonumber
  & \hspace{5cm}\quad \kin{cn, n, s, p, cp},
\end{align}
     where
     \begin{subequations}
       \label{leading_Q}
     \begin{align}
       \asy{Q\ts{Ohm,k}}{0} &=  \sigma\ts{k}\left(\pdv{\asy{\phi\ts{s,k}}{0}}{x}\right)^2 - \asy{i\ts{e,k,1}}{0}
       \pdv{\asy{\phi\ts{e,k}}{0}}{x},&&\kin{n, p}, \\
          \asy{Q\ts{Ohm,s}}{0} &=  - \asy{i\ts{e,s,1}}{0}
       \pdv{\asy{\phi\ts{e,s}}{0}}{x}, && \\
          \asy{Q\ts{Ohm,k}}{0} &=
\sigma\ts{k}'|\nablap\asy{\phi\ts{s,k}}{0}|^2, &&\kin{cn, cp}
     \end{align}
     \end{subequations}
and $\asy{Q\ts{rxn,k}}{0}$ and  $\asy{Q\ts{rev,k}}{0}$  are the leading-order terms in (\ref{eqn:thermal:rxn})-(\ref{eqn:thermal:rev}).
          Integrating across the whole cell from $x = -L\ts{cn}$ to $x = 1 + L\ts{cp}$   gives
\begin{equation}
   \mathcal{C}\ts{th} \sum\ts{k} \left( \rho\ts{k} L\ts{k} \right) \pdv{\asy{T}{0}}{t} = \sum\ts{k} \left( \lambda\ts{k} L\ts{k} \right) \nablap^2 \asy{T}{0} + \mathcal{B}\int_{-L\ts{cn}}^{1 + L\ts{cp}} \asy{Q\ts{k}}{0} \dd{x} + \left[ \lambda\ts{k} \pdv{\asy{T\ts{k}}{2}}{x} \right]^{1+L\ts{cp}}_{-L\ts{cn}}, \label{eq:thermal:averaged}
\end{equation}
where $ \asy{Q\ts{k}}{0}= \asy{Q\ts{Ohm,k}}{0} + \asy{Q\ts{rxn,k}}{0} + \asy{Q\ts{rev,k}}{0}$, and it is understood that the integral of $\asy{Q\ts{k}}{0}$ is the sum of the integrals over each cell component. The final term of \eqref{eq:thermal:averaged} may be evaluated through the use of the boundary condition \eqref{eqn:thermal:newtoncooling}, which gives
\begin{equation}
    \lambda\ts{cn}\pdv{\asy{T\ts{cn}}{2}}{x}\bigg|_{-L\ts{cn}} = \hp\ts{cn}\asy{T}{0}, \qquad  \lambda\ts{cp}\pdv{\asy{T\ts{cp}}{2}}{x}\bigg|_{1+L\ts{cp}} = -\hp\ts{cn}\asy{T}{0},
\end{equation}
where
\[ \hp\ts{cn}(y,z) = \hp(-L\ts{cn},y,z), \quad \hp\ts{cp}(y,z) = \hp(1+L\ts{cp},y,z), \]
are the heat transfer coefficients for the negative and positive current collectors respectively.
Since our choice of nondimensionalisation is such that
\[\sum\limits\ts{k} \rho\ts{k} L\ts{k} = \sum\limits\ts{k} \lambda\ts{k} L\ts{k} = \sum\limits\ts{k} L\ts{k} = L\ts{cn} + 1 + L\ts{cp} = L,\]
say, the governing equation for the leading-order temperature may be written
\begin{subequations}
\label{DFNCC:thermal}
\begin{equation}
    \mathcal{C}\ts{th} \pdv{\asy{T}{0}}{t} =   \nablap^2\asy{T}{0} + \mathcal{B}\asy{\bar{Q}}{0} - \frac{(\hp\ts{cn}+\hp\ts{cp})}{L} \asy{T}{0}, \label{eqn:thermal:leadingtemp}
\end{equation}
where
\begin{equation*}
   \asy{ \bar{Q}}{0} = \frac{1}{L} \int_{-L\ts{cn}}^{1 + L\ts{cp}} \asy{Q\ts{k}}{0} \dd{x}
\end{equation*}
is the $x$-averaged heat source term.  Equation \eqref{eqn:thermal:leadingtemp} is subject to the initial condition
\begin{equation}
\label{ic:nDim:temp}
    \asy{T}{0}(y,z,0) = T_0,
\end{equation}
and the boundary condition
\begin{equation}
    -\nablap {\asy{T}{0}} \cdot \boldsymbol{n} =  \overline{\hpp} \asy{T}{0} \quad (y,z)\in\partial\Omega,
\end{equation}
where
\[ \overline{\hpp} = \frac{1}{L} \int_{-L\ts{cn}}^{1 + L\ts{cp}} \hpp \dd{x}\]
is the $x$-averaged edge heat transfer coefficient.
\end{subequations}

\subsubsection{Summary}
To leading order the reduced model is the two-dimensional pair-potential problem
\begin{subequations}
  \label{eqn:high_summary}
\begin{equation}
    L\ts{cn}\sigma\ts{cn}'\nablap^2 \asy{\phi\ts{s,cn}}{0} = \mathcal{I},
    \qquad
    L\ts{cp}\sigma\ts{cp}'\nablap^2 \asy{\phi\ts{s,cp}}{0} = -\mathcal{I} \qquad \mbox{ in }  \Omega
\end{equation}
with boundary conditions
\begin{equation}
  \asy{\phi\ts{s,cn}}{0} = 0 \quad \mbox{ on } \partial \Omega\ts{tab,cn,$\perp$}, \qquad
\nablap\asy{\phi\ts{s,cn}}{0}\cdot\boldsymbol{n} = 0 \quad \mbox{ on } \partial \Omega\ts{ext,cn,$\perp$}
\end{equation}
\begin{equation}
-\sigma\ts{cp}' \nablap\asy{\phi\ts{s,cp}}{0}\cdot\boldsymbol{n} = \frac{I_{\text{app}}}{A\ts{tab,cp}}  \  \mbox{ on } \partial\Omega\ts{tab,cp,$\perp$},\qquad
\nablap\asy{\phi\ts{s,cp}}{0}\cdot\boldsymbol{n} = 0 \  \mbox{ on } \partial\Omega\ts{ext,cp,$\perp$},
\end{equation}
where $\mathcal{I}$ is the through-cell current given (at each point $(y,z)$) by a one-dimensional DFN model \eqref{eqn:solid:current:np:full-problem}-\eqref{DFNCC:elecchem}, coupled to the two-dimensional thermal problem
\begin{align}
  \mathcal{C}\ts{th} \pdv{\asy{T}{0}}{t} &=   \nablap^2\asy{T}{0} + \mathcal{B}\asy{\bar{Q}}{0} - \frac{(\hp\ts{cn}+\hp\ts{cp})}{L} \asy{T}{0}\qquad \mbox{ in }\Omega,\\
    -\nablap {\asy{T}{0}} \cdot \boldsymbol{n} &=  \overline{\hpp} \asy{T}{0} \quad \mbox{ on }\partial\Omega.
\end{align}
with initial condition $\asy{T}{0} = T_0$, where the heat source is
\begin{align}
  \asy{\bar{Q}}{0} & = \frac{1}{L}\int_0^1\asy{Q\ts{DFN}}{0}\, \dd{x}  + \frac{L\ts{cn}}{L}  \sigma\ts{cn}'|\nablap\asy{\phi\ts{s,cn}}{0}|^2 + \frac{L\ts{cp}}{L}  \sigma\ts{cp}'|\nablap\asy{\phi\ts{s,cp}}{0}|^2,
\end{align}
where $\asy{Q\ts{DFN}}{0} = \asy{Q\ts{Ohm,k}}{0} + \asy{Q\ts{rxn,k}}{0} + \asy{Q\ts{rev,k}}{0}$ ($\kin{n,s,p}$) is the heat source in the one-dimensional DFN model.
\end{subequations}
The dimensional version of these equations is given in \S SM9.1.

\subsection{The very large conductivity limit}\label{sec:very_high_cond}
The model derived in \S\ref{sec:high_cond} is the distinguished limit in which the resistance to current travelling down the current collector is comparable to that to current travelling through the cell. In applications, in order to ensure the whole cell is used uniformly, the current collectors are designed to be thick enough that the potential on them is approximately uniform.
In this section we analyse this situation by considering the sub-limit $\sigma\ts{k}' \gg 1$.

In  \S\ref{sec:high_cond} we also took the edge cooling coefficient to be asymptotically larger than the surface cooling coefficient, so that both effects appeared in the leading-order heat balance.
In this section we weaken the effect of edge cooling by considering the sub-limit $\hpp \ll 1$. We also suppose that the surface cooling coefficients $\hp\ts{k}$ do not vary spatially,  so that the temperature is also approximately uniform.

We will see that with these approximations the model simplifies considerably.
For ease of exposition we quantify the limits by introducing a single small parameter $\ve$ such that $\sigma\ts{k}' = \sigmapp\ts{k}/\ve$, $\hpp = \hppp \ve$ with $\sigmapp\ts{k}$, $\hppp = \mathcal{O}(1)$ as $\ve \rightarrow 0$.
We now expand the leading-order term of  \S\ref{sec:high_cond} in each variable in powers of $\ve$ as
\begin{equation}
\asy{\phi\ts{s,k}}{0} = \asy{\phi\ts{s,k}}{00} + \ve \asy{\phi\ts{s,k}}{01} + \cdots,
\end{equation}
as $\ve\rightarrow 0$. We will retain both the leading term and the first correction in this expansion in $\ve$, while neglecting the first correction in the expansion in $\delta^2$; thus our results are asymptotically accurate providing $\delta^2 \ll \ve$.
After rewriting $\sigma\ts{k}'$ and $\hpp$ \eqref{eqn:high_summary} become
\begin{subequations}
\begin{equation}
    L\ts{cn}\sigmapp\ts{cn}\nablap^2 \asy{\phi\ts{s,cn}}{0} = \ve \mathcal{I},
    \qquad
    L\ts{cp}\sigmapp\ts{cp}\nablap^2 \asy{\phi\ts{s,cp}}{0} = -\ve \mathcal{I} \qquad \mbox{ in }  \Omega
    \label{very_high_phi_cp}
\end{equation}
\begin{equation}
  \asy{\phi\ts{s,cn}}{0} = 0 \quad \mbox{ on } \partial \Omega\ts{tab,cn,$\perp$}, \qquad
\nablap\asy{\phi\ts{s,cn}}{0}\cdot\boldsymbol{n} = 0 \quad \mbox{ on } \partial \Omega\ts{ext,cn,$\perp$}
\end{equation}
\begin{equation}
-\sigmapp\ts{cp} \nablap\asy{\phi\ts{s,cp}}{0}\cdot\boldsymbol{n} = \frac{\ve I_{\text{app}}}{A\ts{tab,cp}}  \  \mbox{ on } \partial\Omega\ts{tab,cp,$\perp$},\qquad
\nablap\asy{\phi\ts{s,cp}}{0}\cdot\boldsymbol{n} = 0 \
\mbox{ on } \partial\Omega\ts{ext,cp,$\perp$},\label{very_high_phi_cp_bc}
\end{equation}
\begin{align}
  \mathcal{C}\ts{th} \pdv{\asy{T}{0}}{t} &=   \nablap^2\asy{T}{0} + \mathcal{B}\asy{\bar{Q}}{0} - \frac{(\hp\ts{cn}+\hp\ts{cp})}{L} \asy{T}{0}\qquad \mbox{ in }\Omega,\label{very_high_T}\\
    -\nablap {\asy{T}{0}} \cdot \boldsymbol{n} &= \ve \overline{\hppp} \asy{T}{0} \quad \mbox{ on }\partial\Omega.\label{very_high_T_bc}
\end{align}
with initial condition $\asy{T}{0} = T_0$. It is useful to write down also a global current conservation equation, by integrating the second equation in \eqref{very_high_phi_cp} over $\Omega$ and using \eqref{very_high_phi_cp_bc} to give
\begin{equation}
    I\ts{app} =  \int_{\Omega} \mathcal{I} \dd{y}\dd{z}\label{very_high_global_current}.
\end{equation}

\end{subequations}

\subsubsection{Leading-order problem}
At leading order in $\ve$ we find the potentials are uniform as expected, with
\begin{equation}
    \asy{\phi\ts{s,cn}}{00} = 0, \qquad \asy{\phi\ts{s,cp}}{00} = \asy{V}{00}(t),
\end{equation}
where $\asy{V}{00}(t)$ is the (unknown) leading-order terminal voltage.
Since \eqref{very_high_T_bc} gives
\[   -\nablap {\asy{T}{00}} \cdot \boldsymbol{n} = 0 \quad \mbox{ on }\partial\Omega,\]
the leading-order temperature $\asy{T}{00}$ will be spatially uniform if the heat source  $\asy{\bar{Q}}{00}$ is spatially uniform. On the other hand, if $\asy{T}{00}$ is spatially uniform then at each point $(y,z)$  the one-dimensional DFN
model \eqref{DFNCC:eleclyte}-\eqref{DFNCC:elecchem} sees the same temperature $\asy{T}{00}$ and potential difference $V(t)$, so that (providing the initial condition is independent of $y$ and $z$)  the solution to each of these models is independent of $y$ and $z$, and the through-cell current $\asy{\mathcal{I}}{00}$ and heating $\asy{\bar{Q}}{00}$ are uniform. Thus a  single one-dimensional DFN
problem  suffices to determine $\asy{V}{00}$ as a functional of $\asy{\mathcal{I}}{00}$ and $\asy{T}{00}$. Let us write this output of the DFN model as
\begin{equation}
  \phi\ts{s,cp} - \phi\ts{s,cn} = V\ts{DFN}(\mathcal{I},T),\label{DFN_soln}
\end{equation}
so that $\asy{V}{00} = V\ts{DFN}(\asy{\mathcal{I}}{00},\asy{T}{00})$.
The current $\asy{\mathcal{I}}{00}$ is given by \eqref{very_high_global_current} as
\begin{equation}
    \asy{\mathcal{I}}{00} =\frac{ I\ts{app}}{L_y L_z},
\end{equation}
while the leading-order temperature is determined from  the ordinary differential equation
\begin{subequations}\label{eqn:high-cc:leading-order-thermal}
\begin{gather}
\mathcal{C}\ts{th} \dv{\asy{T}{00}}{t} =  \mathcal{B} \bar{Q}\ts{DFN}(\asy{\mathcal{I}}{00},\asy{T}{00})  - \frac{(\hp\ts{cn}+\hp\ts{cp})}{L} \asy{T}{00},
\end{gather}
\end{subequations}
with initial condition $\asy{T}{00}(0) = T_0$,
where
\begin{align}
  \bar{Q}\ts{DFN}(\mathcal{I},T) & = \frac{1}{L}\int_0^1\asy{Q\ts{DFN}}{0}\, \dd{x},
\end{align}
is the $x$-averaged heat source in the one-dimensional DFN model, which, like $V\ts{DFN}$ is  a functional of the  temperature $T$ and current $\mathcal{I}$.

\subsubsection{First-order correction}
The  first-order corrections to the current collector potentials satisfy
\begin{subequations}
\begin{equation}
    L\ts{cn}\sigmapp\ts{cn}\nablap^2 \asy{\phi\ts{s,cn}}{01} =   \frac{I\ts{app}}{L_y L_z},
    \qquad
    L\ts{cp}\sigmapp\ts{cp}\nablap^2 \asy{\phi\ts{s,cp}}{01} = - \frac{I\ts{app}}{L_y L_z} \qquad \mbox{ in }  \Omega
    \label{very_high_phi_cp_first}
\end{equation}
\begin{equation}
  \asy{\phi\ts{s,cn}}{01} = 0 \quad \mbox{ on } \partial \Omega\ts{tab,cn,$\perp$}, \qquad
\nablap\asy{\phi\ts{s,cn}}{01}\cdot\boldsymbol{n} = 0 \quad \mbox{ on } \partial \Omega\ts{ext,cn,$\perp$}
\end{equation}
\begin{equation}
-\sigmapp\ts{cp} \nablap\asy{\phi\ts{s,cp}}{01}\cdot\boldsymbol{n} = \frac{I_{\text{app}}}{A\ts{tab,cp}}  \  \mbox{ on } \partial\Omega\ts{tab,cp,$\perp$},\qquad
\nablap\asy{\phi\ts{s,cp}}{01}\cdot\boldsymbol{n} = 0 \
\mbox{ on } \partial\Omega\ts{ext,cp,$\perp$},\label{very_high_phi_cp_bc_first}
\end{equation}
\end{subequations}
We note that $\asy{\phi\ts{s,cp}}{01}$ is only determined up to a function of time,  which is fixed by solving the through-cell DFN problem at ${\mathcal O}(\ve)$. However, we will see that we can evaluate this term without solving multiple DFN models parameterised by $y$ and $z$.
We  use \eqref{DFN_soln} to write
\[ \asy{\phi\ts{s,cp}}{01} - \asy{\phi\ts{s,cn}}{01} = \frac{\delta V\ts{DFN}}{\delta \mathcal{I}} \asy{\mathcal{I}}{01} +
\frac{\delta V\ts{DFN}}{\delta T} \asy{T}{01}\]
where the functional derivatives are evaluated at $(\asy{\mathcal{I}}{00},\asy{T}{00})$, and are therefore independent of $y$ and $z$.
Integrating over $\Omega$ gives
\begin{equation}
  \av{\asy{\phi\ts{s,cp}}{01}} - \av{\asy{\phi\ts{s,cn}}{01}} = \frac{\delta V\ts{DFN}}{\delta \mathcal{I}} \av{\asy{\mathcal{I}}{01}} +
  \frac{\delta V\ts{DFN}}{\delta T} \av{\asy{T}{01}}
  \label{DFNeps}
\end{equation}
where
\[ \av{\cdot} = \frac{1}{L_y L_z} \int_{\Omega} \cdot \, \dd{y}\, \dd{z}.\]
But (\ref{very_high_global_current}) gives
\[  \av{\asy{\mathcal{I}}{01}} = 0,\]
so the only contribution from the DFN at $O(\ve)$ is from the temperature perturbation.

Recalling that the terminal voltage is the average of the potential over the positive tab,
\[ \asy{V}{01} = \frac{L\ts{cp}}{A\ts{tab,cp}}\int_{\partial\Omega\ts{tab,cp,$\perp$}}  \asy{\phi\ts{s,cp}}{01}\, \dd{s},\]
and \eqref{very_high_phi_cp_first} is enough to determine
$\av{\asy{\phi\ts{s,cn}}{01}}$ and $\asy{V}{01}-\av{\asy{\phi\ts{s,cp}}{01}}$, which can be interpreted as the potential drops across the negative and positive current collectors respectively. Since these are proportional to $I\ts{app}$ (which may be time dependent) they can be most easily formulated in terms of current collector resistances by writing
\begin{equation}
  \av{\asy{\phi\ts{s,cn}}{01}} = - R\ts{cn} I\ts{app}, \qquad
  \asy{V}{01}-\av{\asy{\phi\ts{s,cp}}{01}} = - R\ts{cp} I\ts{app},
  \label{cc_resistances}
\end{equation}
where
\begin{equation} R\ts{cn} = \frac{ \av{f\ts{n}}}{L_y L_z L\ts{cn}\sigmapp\ts{cn}}, \qquad R\ts{cp} =  \frac{1}{L_y L_z \sigmapp\ts{cp}A\ts{tab,cp}}\int_{\partial\Omega\ts{tab,cp,$\perp$}}   f\ts{p}\, \dd{s}
\end{equation}
with
\begin{subequations}
\begin{equation}
    \nablap^2 f\ts{n}=  -1, \qquad  \qquad
   \nablap^2 f\ts{p} = 1 \qquad \mbox{ in }  \Omega,
\end{equation}
\begin{align}
  f\ts{n} &= 0 \quad \mbox{ on } \partial \Omega\ts{tab,cn,$\perp$},&
\nablap f\ts{n}\cdot\boldsymbol{n} &= 0 \quad \mbox{ on } \partial \Omega\ts{ext,cn,$\perp$}  ,\\
\nablap f\ts{p}\cdot\boldsymbol{n} &= \frac{L_y L_zL\ts{cp}}{A\ts{tab,cp}}  \  \mbox{ on } \partial\Omega\ts{tab,cp,$\perp$},&
\nablap f\ts{p}\cdot\boldsymbol{n} &= 0 \
\mbox{ on } \partial\Omega\ts{ext,cp,$\perp$}, \qquad \av{f\ts{p}}=0.
\end{align}
\end{subequations}
Combining \eqref{cc_resistances} with \eqref{DFNeps} gives the perturbation to the terminal voltage as
\begin{equation}
\label{eqn:V01}  \asy{V}{01} =
  \frac{\delta V\ts{DFN}}{\delta T} \av{\asy{T}{01}}-R\ts{cp} I\ts{app}-R\ts{cn} I\ts{app}.
\end{equation}
At next order in \eqref{very_high_T}-\eqref{very_high_T_bc} we find
\begin{align*}
  \mathcal{C}\ts{th} \pdv{\asy{T}{01}}{t} &=   \nablap^2\asy{T}{01} + \mathcal{B}\frac{\delta \bar{Q}\ts{DFN}}{\delta T}\asy{T}{01} - \frac{(\hp\ts{cn}+\hp\ts{cp})}{L} \asy{T}{01} \\
  & \qquad \qquad \mbox{ }+ \frac{\mathcal{B}L\ts{cn}\sigmapp\ts{cn}}{L}  |\nablap\asy{\phi\ts{s,cn}}{01}|^2+ \frac{\mathcal{B}L\ts{cp}\sigmapp\ts{cp}}{L}  |\nablap\asy{\phi\ts{s,cp}}{01}|^2\qquad \mbox{ in }\Omega,\\
    -\nablap {\asy{T}{01}} \cdot \boldsymbol{n} &=  \overline{\hppp} \asy{T}{00} \quad \mbox{ on }\partial\Omega,
\end{align*}
where the functional derivative is evaluated at $\asy{T}{00}$.
Integrating over $y$ and $z$ gives
\begin{align}
\label{eqn:avT01}  \mathcal{C}\ts{th} \pdv{\av{\asy{T}{01}}}{t} &=    \mathcal{B}\frac{\delta \bar{Q}\ts{DFN}}{\delta T}\av{\asy{T}{01}} - \frac{(\hp\ts{cn}+\hp\ts{cp})}{L} \av{\asy{T}{01}}\\ \nonumber
  & \hspace{2cm} \mbox{ }- \frac{\asy{T}{00}}{L_y L_z} \int_{\partial \Omega}\overline{\hppp}\, \dd{s} + H\ts{cn} I\ts{app}^2+H\ts{cp} I\ts{app}^2  \qquad \mbox{ in }\Omega,
\end{align}
where the coefficients related to Ohmic heating in the current collectors are
\begin{equation}
H\ts{cn} = \frac{\mathcal{B}L\ts{cn}}{L(L_y L_z L\ts{cn})^2\sigmapp\ts{cn}} \av{|\nablap f\ts{n}|^2}, \qquad H\ts{cp} = \frac{\mathcal{B}L\ts{cp}}{L(L_y L_z L\ts{cp})^2\sigmapp\ts{cp}}  \av{|\nablap f\ts{p}|^2}.
\end{equation}
In principle (\ref{eqn:avT01}) allows the correction to the average temperature to be determined, whence \eqref{eqn:V01} gives the correction to the terminal voltage.
Rather than evaluating $\delta V\ts{DFN}/\delta T$ and $\delta \bar{Q}\ts{DFN}/\delta T$ the most convenient way to capture the perturbation is to note that
\begin{align*}
  V\ts{DFN}(\asy{\mathcal{I}}{00}, \asy{T}{00}+\ve \av{\asy{T}{01}}) &=
  V\ts{DFN}(\asy{\mathcal{I}}{00}, \asy{T}{00})\\
  & \qquad \mbox{ }+ \ve\frac{\delta V\ts{DFN}}{\delta T}
(\asy{\mathcal{I}}{00}, \asy{T}{00})
  \av{\asy{T}{01}} + O(\ve^2),
  \end{align*}
so that
\[ \asy{V}{00} + \ve \asy{V}{01} =
  V\ts{DFN}(\asy{\mathcal{I}}{00}, \asy{T}{00}+\ve \av{\asy{T}{01}})-\ve R\ts{cp} I\ts{app}-\ve R\ts{cn} I\ts{app}+ O(\ve^2). \]
  Thus   we may solve a single one-dimensional DFN using the $y,z$-averaged temperature, and the error will be $O(\ve^2)$.

  \subsubsection{Summary}
  Writing $\av{T} = \asy{T}{00} + \ve \av{\asy{T}{01}}$ gives
  \begin{subequations}
    \label{model_very_high}
    \begin{align}
  \label{eq:very_high_V}
  V & =  V\ts{DFN}(\asy{\mathcal{I}}{00}, \av{T})-\ve R\ts{cp} I\ts{app}-\ve R\ts{cn} I\ts{app}+ O(\ve^2, \delta^2),\\
  \label{eq:very_high_T}
\mathcal{C}\ts{th} \pdv{\av{T}}{t} &=    \mathcal{B}\bar{Q}\ts{DFN}(\asy{\mathcal{I}}{00}, \av{T}) - \frac{(\hp\ts{cn}+\hp\ts{cp})}{L} \av{T}- \frac{\ve \av{T}}{L_y L_z} \int_{\partial \Omega}\overline{\hppp}\, \dd{s} \\ \nonumber
  & \hspace{3cm} \mbox{ }+ \ve H\ts{cn} I\ts{app}^2+\ve H\ts{cp} I\ts{app}^2 + O(\ve^2, \delta^2) \qquad \mbox{ in }\Omega.
\end{align}
After solving this single one-dimensional model,  the potential distribution in the current collectors is
\begin{equation}
  \label{eq:very_high_phi_cc}
  \phi\ts{s,cn} = -\frac{\ve\asy{\mathcal{I}}{00}}{L\ts{cn}\sigmapp\ts{cn}} f\ts{n}+O(\ve^2, \delta^2), \qquad
\phi\ts{s,cp} = V + \frac{\ve\asy{\mathcal{I}}{00}}{L\ts{cp}\sigmapp\ts{cp}} f\ts{p}+O(\ve^2, \delta^2).
\end{equation}
Recall that in this limit the leading-order current is $\asy{\mathcal{I}}{00} = I\ts{app}/(L_y L_z).$
\end{subequations}
The dimensional version of these equations is given in \S SM9.1.1.
\subsubsection{An ad-hoc model for the temperature distribution}
The reduced model (\ref{model_very_high})
gives the spatial variation of the potential in the current collectors, but only the average cell temperature. An approach sometimes used in the literature is to retain the spatial derivatives in the energy balance equation, but use heat source terms from the averaged one-dimensional electrochemical model (see e.g.~\cite{hosseinzadeh2018}). Such an approach corresponds to replacing
$\bar{Q}\ts{DFN}(\asy{\mathcal{I}}{00}, T)$ with $\bar{Q}\ts{DFN}(\asy{\mathcal{I}}{00}, \av{T})$
and
replaces \eqref{eq:very_high_T} with
\begin{subequations}
\begin{align}
\mathcal{C}\ts{th} \pdv{T}{t} &=   \nablap^2 T+ \mathcal{B}\bar{Q}\ts{DFN}(\asy{\mathcal{I}}{00}, \av{T}) - \frac{(\hp\ts{cn}+\hp\ts{cp})}{L} T \\ \nonumber
& \hspace{1cm} \mbox{ } + \frac{\mathcal{B}L\ts{cn}\sigmapp\ts{cn}}{\ve L}  |\nablap\phi\ts{s,cn}|^2+ \frac{\mathcal{B}L\ts{cp}\sigmapp\ts{cp}}{\ve L}  |\nablap\phi\ts{s,cp}|^2
\qquad \mbox{ in }\Omega,\\
-\nablap T \cdot \boldsymbol{n} &=  \ve\overline{\hppp} T \quad \mbox{ on }\partial\Omega.
\end{align}
\end{subequations}
This model captures the variation due to Ohmic heating in the current collectors and  cooling at the boundaries, but neglects the spatial variation of the  heat source within the cell.

\section{Comparison of Models}\label{sec:results}

In this section, we provide a numerical comparison of the full model with the reduced models \eqref{eqn:high_summary} and \eqref{model_very_high}.
For ease of exposition we focus  on  the case in which all variables are uniform in $y$, so that the full model is two-dimensional, and the reduced model \eqref{eqn:high_summary} has a one-dimensional current collector (in the $z$ direction), at each point of which we solve a one-dimenisonal DFN model (we refer to this as a 1+1D model, though since the DFN is already a pseudo-two-dimensional model, perhaps it is more properly a 1+1+1D model). We refer to the very-high conductivity limit model \eqref{model_very_high} as the DFNCC model, to indicate that it involves a single (averaged) DFN model with an  additional (uncoupled) problem for the distribution of potential in the current collectors (from which the resistance and heat source can be calculated).

Numerical simulations of the full model were performed using the commercial software COMSOL \cite{COMSOL}, while the reduced models were implemented in the open-source battery modelling package PyBaMM (Python Battery Mathematical Modelling) \cite{pybamm}.
All simulations were performed on a desktop computer (i5, 2.1 GHz) with 16 Gb of RAM. The model equations in COMSOL are discretised in space using the finite element method, while in PyBaMM  the equations are discretised using the finite volume method.
Both solvers use an adaptive, variable-order backward differentiation formula for the time integration, with both relative and absolute tolerances set to $10^{-6}$. Since we aim to compare the full and reduced models and not the merits of any particular numerical  approach, we provide a comparison of the solutions of the standard one-dimensional DFN model produced by both COMSOL and PyBaMM in the supplementary material. Whilst this does not fully quantify differences introduced by employing two different numerical solution methods, it does provide context for our comparisons that  follow.

Typical dimensional parameter values for battery comprising a carbon negative current collector, graphite negative electrode, LiPF$_6$ in EC:DMC
electrolyte, LCO positive electrode, and aluminium positive current collector  are given in Table~SM1 (taken from \cite{SMouraGithub}). These  translate into the nondimensional parameters given in Table \ref{table:dimensionless_parameter_values}. From there we see that
\[ h \approx 3.8 \times 10^{-5}, \quad \delta \approx 1.6 \times 10^{-3}.\]
The dimensionless conductivities $\sigma\ts{cn}$ and $\sigma\ts{cp}$ depend on the charge/discharge rate (the so-called C-rate)\footnote{It is standard practice in the field to measure C-rates in multiples of the rate at which the battery would charge/discharge in 1 hour, known as 1C. Thus, for example,  at a discharge rate 2C the battery would discharge in 30 minutes.}.
To give an idea of the typical asymptotic regime batteries operate in, at a C-rate of 3 we find that
\begin{equation}
    \sigma\ts{cn} \approx 9.5 \times 10^7, \quad \sigma\ts{cp} \approx 5.6 \times 10^7.
\end{equation}

\begin{figure}
    \centering
    \includegraphics[width=\textwidth]{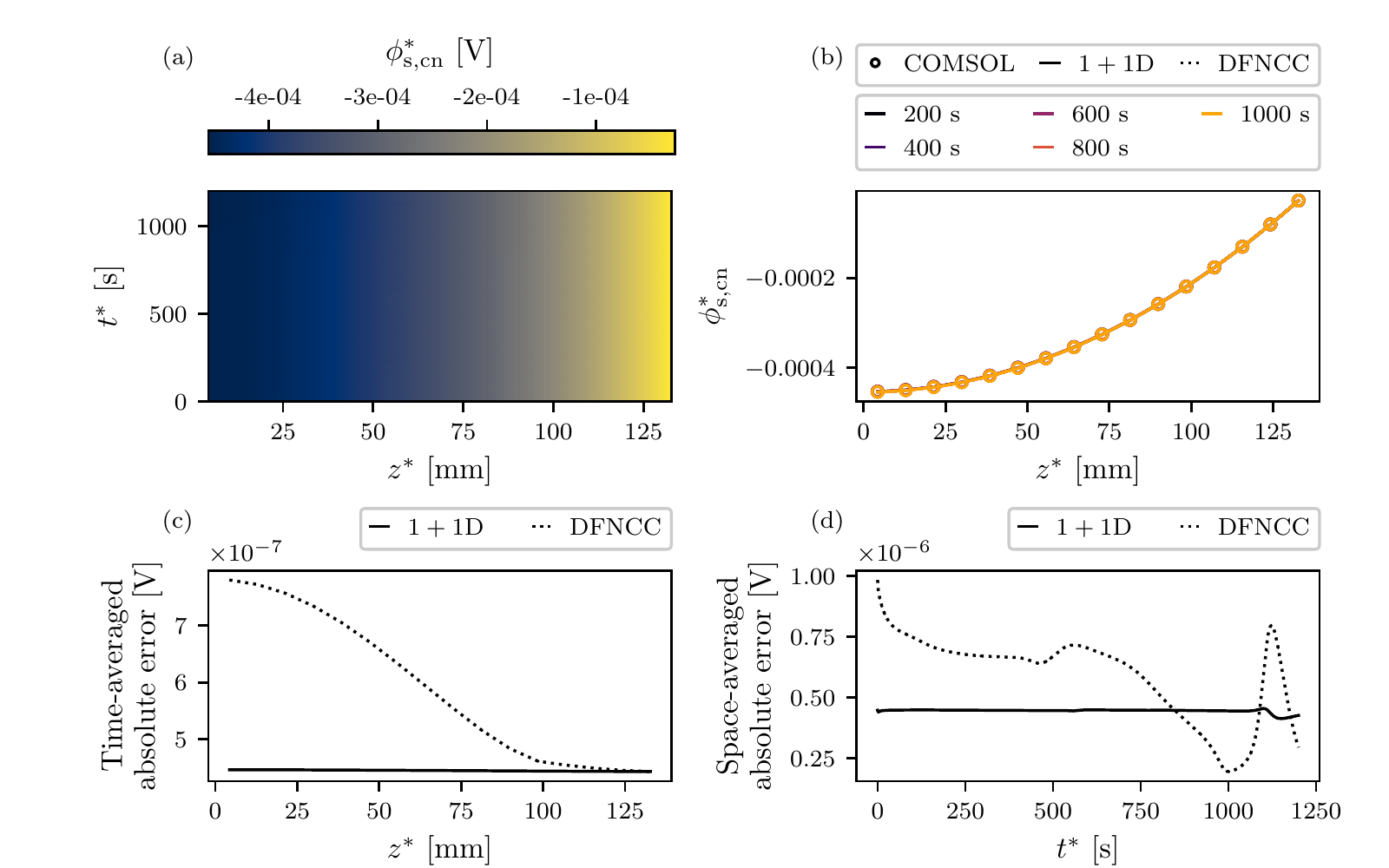}
    \caption{
      Potential in the negative current collector.
      (a) the COMSOL solution; (b) comparison with the reduced  models at various times during  discharge; (c) time-averaged absolute errors; (d) $z$-averaged absolute errors.
          }
    \label{fig:1plus1D_phi_cn_comparison}
\end{figure}

\begin{figure}
    \centering
    \includegraphics[width=\textwidth]{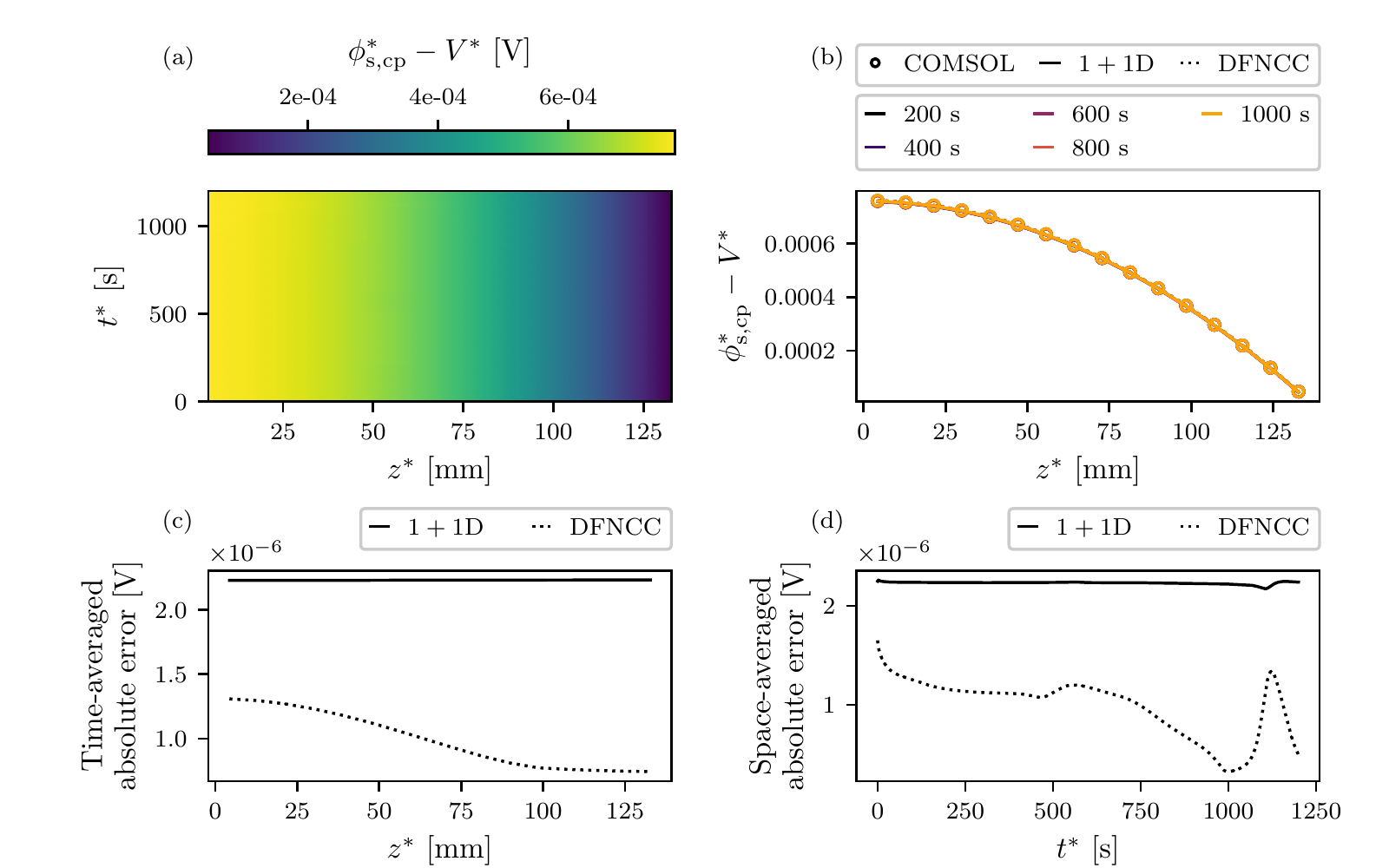}
    \caption{
Potential in the positive current collector (with the terminal voltage subtracted off).
      (a) the COMSOL solution; (b) comparison with the reduced  models at various times during  discharge; (c) time-averaged absolute errors; (d) $z$-averaged absolute errors.
    }
    \label{fig:1plus1D_phi_cp_comparison}
\end{figure}

\begin{figure}
    \centering
    \includegraphics[width=\textwidth]{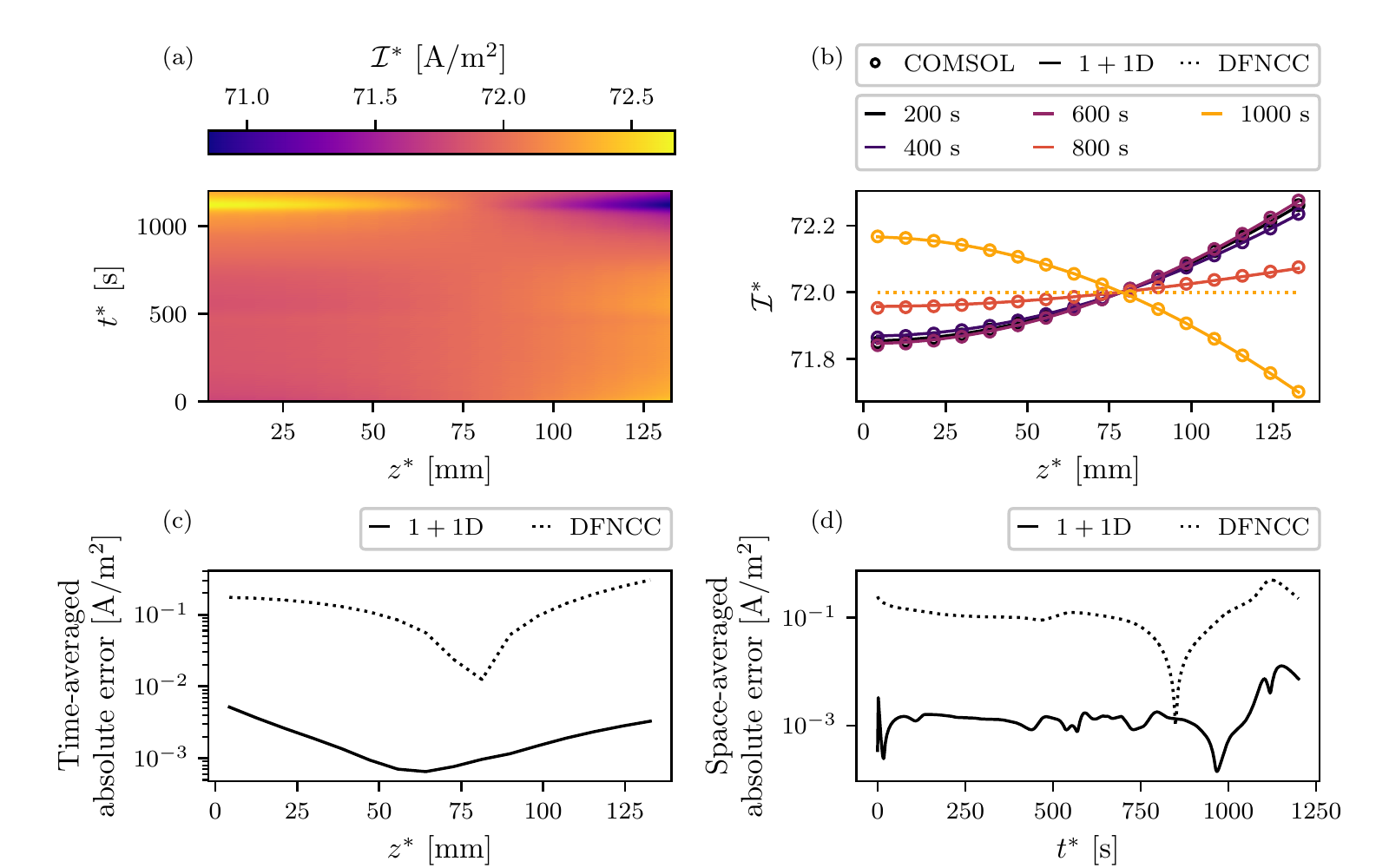}
    \caption{
      Through-cell current density.
      (a) the COMSOL solution; (b) comparison with the reduced  models at various times during  discharge; (c) time-averaged absolute errors; (d) $z$-averaged absolute errors.
     }
    \label{fig:1plus1D_I_comparison}
\end{figure}

\begin{figure}
    \centering
    \includegraphics[width=\textwidth]{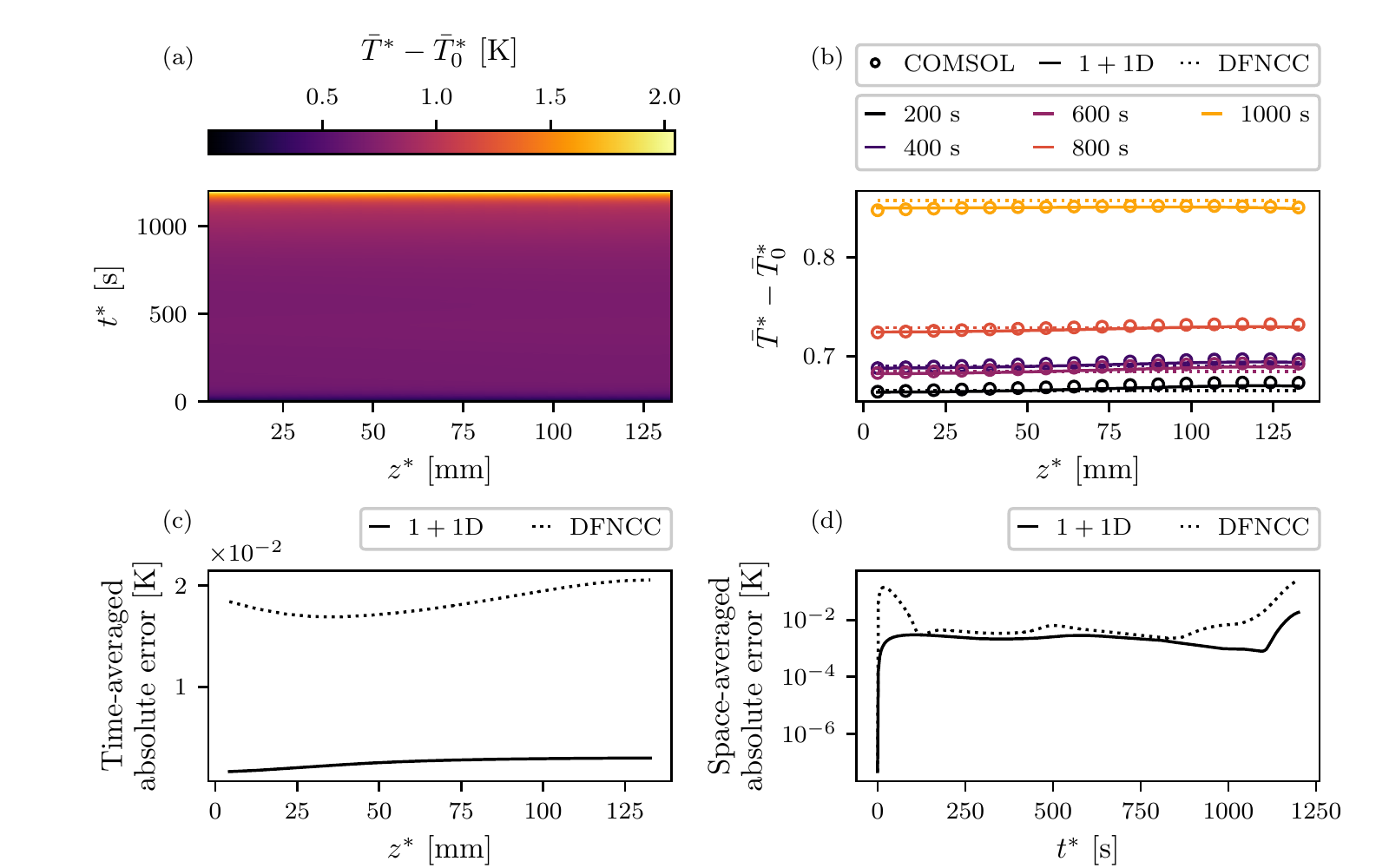}
    \caption{
      The $x$-averaged temperature
      (a) the COMSOL solution; (b) comparison with the reduced  models at various times during  discharge; (c) time-averaged absolute errors; (d) $z$-averaged absolute errors.
}
    \label{fig:1plus1D_T_comparison}
\end{figure}

We compare the results of the 2D DFN model with the 1+1D DFN and DFNCC models for a 3C
 constant current discharge, with both positive and negative tabs placed at the  top of the cell (i.e. at $z = L_z$). In Figures~\ref{fig:1plus1D_phi_cn_comparison}, \ref{fig:1plus1D_phi_cp_comparison}, \ref{fig:1plus1D_I_comparison} and \ref{fig:1plus1D_T_comparison} we present comparisons for the potential in the negative current collector, the potential in the positive current collector, the through-cell current density and the $x$-averaged temperature respectively.
Solutions from the full model are shown as a function of space in time in panel (a), with snapshots at a series of times throughout the discharge shown in panel (b). The time- and space-averaged absolute errors\footnote{By ``error'' we mean the  difference between the numerical solution of the reduced model in PyBaMM and the COMSOL solution of the full model.} are shown in panels (c) and (d), respectively.

We see in Figures~\ref{fig:1plus1D_phi_cn_comparison} and \ref{fig:1plus1D_phi_cp_comparison} that the electrical conductivity of the current collectors is sufficiently high that the potentials remain fairly uniform in space, and both the 1+1D DFN and DFNCC models are able to accurately capture the potential distribution in the current collectors.
The error  is of a similar size to that between the numerical solutions of the 1D DFN obtained using COMSOL and PyBaMM (\S SM8), so that little additional error has been introduced as a result of the asymptotic reduction.

In Figure~\ref{fig:1plus1D_I_comparison}(a) we see that  positioning  both tabs  at the top of the cell means that for most of the simulation the current preferentially travels through the upper part of the cell. Eventually, as the cell continues to discharge, this part becomes more (de)lithiated until the resultant local increase in through-cell resistance  is sufficient for it to become preferential for the current to travel further along the current collectors and through the lower part of the cell (as seen in the final time shown in Figure~\ref{fig:1plus1D_I_comparison}(b)). This behaviour is well captured by the 1+1D model, with space-averaged absolute errors in the through-cell current on the order of $10^{-3}$\si{A.m^{-2}} for most of the discharge, as displayed in Figure~\ref{fig:1plus1D_I_comparison}(d). The largest error is found towards the end of the discharge where the OCV becomes highly nonlinear\footnote{Although this is also where the greatest discrepancy in the solution between COMSOL and PyBaMM in 1D is found (see the 1C result in Figure~SM1).}. In the DFNCC formulation the through-cell current density is assumed uniform, so the greatest error is found at the ends of the current collectors where the current density deviates most from its average.

For the parameters given in Table~SM1 we find that the temperature exhibits a relatively weak variation along the length of the current collectors, as shown in Figure~\ref{fig:1plus1D_T_comparison}.
The 1+1D model captures the temperature distribution well.

Since the temperature rise is moderate (and the variation of temperature in space is small), the uniform temperature predicted by the DFNCC model gives a good estimate of the temperature in the full model.

In Table~\ref{table:1plus1D_errors_vs_npts}, we give the normalised root mean square (RMS) error in the current collector potentials, through-cell current, temperature, and voltage obtained by solving the model in PyBaMM as the mesh is refined. The RMS error was computed with respect to the solution obtained using COMSOL's ``fine'' mesh (450 elements in each current collector, 1650 elements in each electrode, 450 elements in the separator), which was typically solved in around 5376s.
It can be seen that the error in the through-cell current density $\mathcal{I}^*$ for the DFNCC model is much larger than that of the $1+1$D model, and remains unchanged as the mesh is refined: this is the asymptotic error inherent in the model.
 However, other quantities, such as the terminal voltage, are predicted equally well by the DFNCC model at a fraction of the computation time. Depending on the quantities of interest the simpler DFNCC model may well be sufficient for a range of applications.

\begin{figure}
    \centering
    \includegraphics[width=\textwidth]{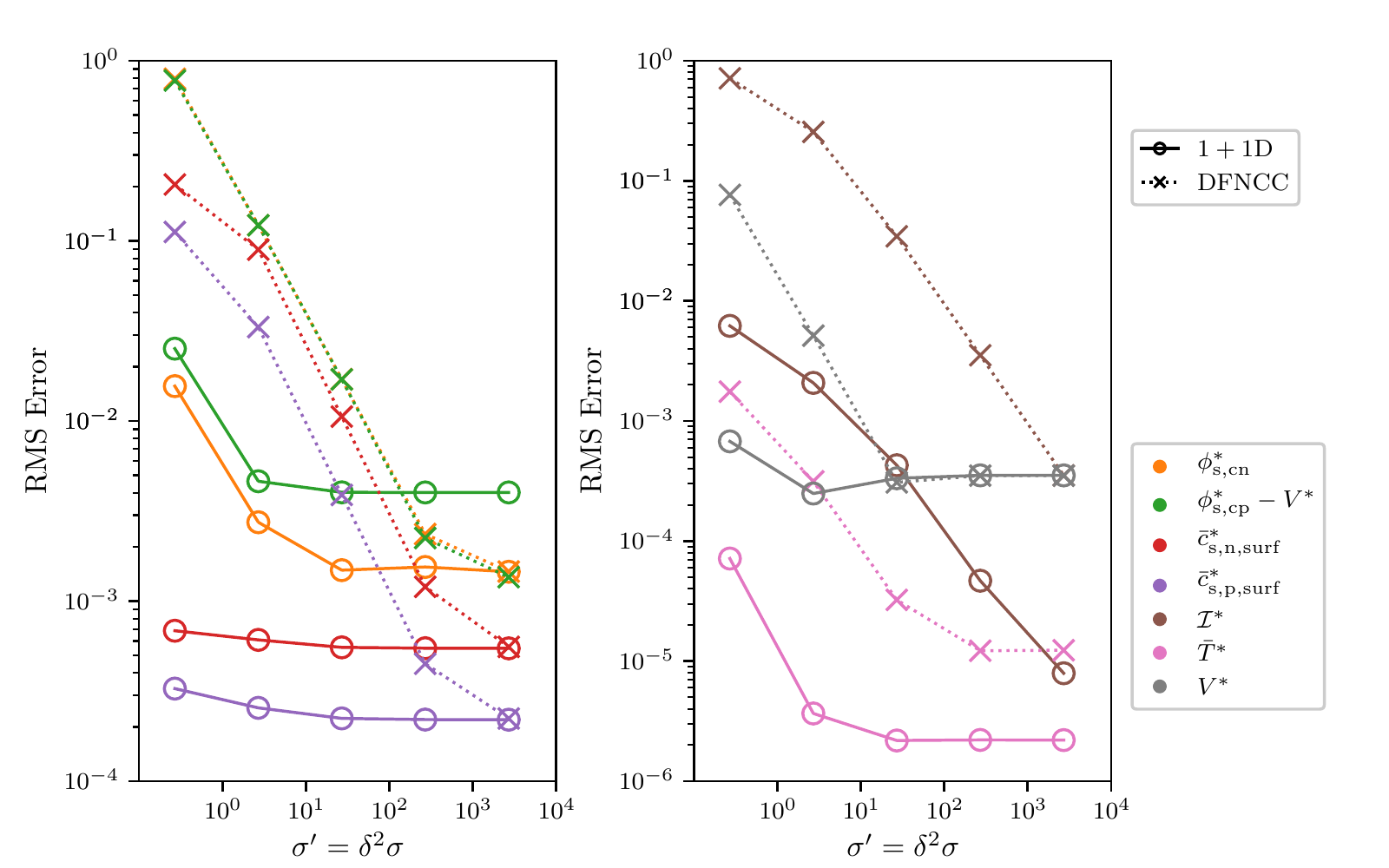}
    \caption{Normalised RMS error between the 2D solution in COMSOL and the 1+1D DFN and DFNCC solutions in PyBaMM for a selection of model variables as the non-dimensional conductivity $\sigma$ is varied, with $\sigma\ts{cn}=\sigma\ts{cp}=\sigma$. The quantities plotted for a variable $\psi$ were computed as RMS($\psi\ts{PyBaMM}-\psi\ts{COMSOL}$)/RMS($\psi\ts{COMSOL}$).
}
    \label{fig:RMSE_sigma}
\end{figure}

 Finally, to illustrate the asymptotic convergence of the DFNCC model we
 fix $\sigma\ts{cn}=\sigma\ts{cp}=\sigma$ and solve for a range of values of $\sigma$. The normalised RMS error between the 2D solution in COMSOL and the 1+1D DFN and DFNCC solutions in PyBaMM for a selection of model variables are shown in Figure~\ref{fig:RMSE_sigma}.

\begin{table}[htb]
	\centering
    \resizebox{\textwidth}{!}{
	\begin{tabular}{c c c c c c c c c}
	\toprule
\multicolumn{9}{c}{1+1D}\\
\toprule
        $N$ & $\phi\ts{s,cn}^*$& $\phi\ts{s,cp}^* - V^*$ &  $\bar{c}\ts{s,n,surf}^*$  & $\bar{c}\ts{s,p,surf}^*$ & $\mathcal{I}^*$ & $\bar{T}^*$ & $V^*$ &  Solution time [s] \\
    \midrule
        4 & \num{2.148e-2}  & \num{6.420e-2} & \num{1.646e-2} & \num{2.676e-3} & \num{3.954e-4} & \num{1.024e-4} & \num{2.341e-3} & 0.6115  \\
        8 & \num{5.377e-3}  & \num{1.605e-2} & \num{5.767e-3} & \num{7.584e-4} & \num{1.015e-4} & \num{3.030e-5} & \num{6.864e-4} &  1.323\\
        16 & \num{1.345e-3} & \num{4.011e-3} & \num{1.815e-3} & \num{2.082e-4} & \num{3.249e-5} & \num{7.772e-6} & \num{1.774e-4} & 9.446\\
        32 & \num{3.421e-4} & \num{1.004e-3} & \num{5.231e-4} & \num{5.459e-5} & \num{2.665e-5} & \num{3.056e-6} & \num{4.412e-5} &  85.97\\
    \bottomrule
	\toprule
\multicolumn{9}{c}{DFNCC}\\
	\toprule
     $N$ & $\phi\ts{s,cn}^*$& $\phi\ts{s,cp}^* - V^*$ &  $\bar{c}\ts{s,n,surf}^*$  & $\bar{c}\ts{s,p,surf}^*$ & $\mathcal{I}^*$ & $\bar{T}^*$ & $V^*$ &  Solution time [s] \\
    \midrule
     4 & \num{2.172e-2}& \num{2.172e-2}& \num{1.650e-2} & \num{2.703e-3} & \num{2.120e-3}& \num{1.202e-4} & \num{2.339e-3} & 0.24 \\
     8 & \num{5.725e-3} & \num{5.725e-3} & \num{5.895e-3} & \num{8.527e-4} & \num{2.294e-3} &\num{8.622e-5} & \num{6.801e-4} & 0.41 \\
     16 & \num{1.948e-3}&  \num{1.948e-3} &  \num{2.194e-3} & \num{4.437e-4} & \num{2.330e-3} & \num{8.612e-5} & \num{1.728e-4} & 0.99 \\
     32 & \num{1.262e-3} & \num{1.262e-3} &  \num{1.241e-3} & \num{3.963e-4} & \num{2.334e-3} &  \num{8.724e-5} & \num{4.931e-5} & 2.9 \\
    \bottomrule
	\end{tabular}
	} 
    \caption{Normalised RMS error between the 2D solution in COMSOL and the 1+1D DFN and DFNCC solutions in PyBaMM for a selection of model variables. The tabulated quantities for a variable $\psi$ were computed as RMS($\psi\ts{PyBaMM}-\psi\ts{COMSOL}$)/RMS($\psi\ts{COMSOL}$). Here $N$ is the number of mesh cells per spatial dimension in each domain in the 1+1D  model. The 1+1D  solution was compared to the 2D  solution on a ``fine'' mesh (450 elements in each current collector, 1650 elements in each electrode, 450 elements in the separator) in COMSOL. Both time stepping routines used a relative and absolute tolerance of $10^{-6}$.
    }\label{table:1plus1D_errors_vs_npts}
\end{table}

\section{Conclusions}\label{sec:conclusion}
In this paper, we have provided a systematic asymptotic derivation of the 2+1D DFN  battery model from the full 3D DFN model, identifying the key non-dimensional parameters controling the reduction.
Our findings are in agreement with other works that employ the 2+1D approach in an ad-hoc fashion (e.g. \cite{kosch2018, gerver2011, lee2013, kim2011}).  Moreover, we have shown that, in a suitable parameter regime, the model can be simplified further to the DFNCC comprising a single representative 1D model describing the electrochemistry in the through-cell direction with an uncoupled two dimensional problem to solve for the distribution of potential in the current collectors, from which resistances and heat generation can be determined. This latter approximation reduces the model from pseudo-four-dimensional  to pseudo-two-dimensional, dramatically reducing computational cost.

By identifying the parameters which control the asymptotic reduction our analysis highlights the parameter regimes in which the 2+1D DFN and DFNCC models are appropriate, and quantifies the error a priori.
This in turn informs practical design choices for key cell parameters
(such as current collector thickness or tab placement) in order that the cell discharge uniformly.

Our systematic analysis also makes clear that the simplifications are independent of the model used for the through-cell current, so that they can be combined with other through-cell asymptotic simplifications (e.g. reducing the DFN to the SPMe \cite{Marquis2019, richardson2019generalised}) in a systematic and mathematically consistent way. This idea is exploited in \cite{sulzer_part_II}, in which further model reductions are considered in various interesting and physically relevant limits.

Of course, more complicated through-cell models can also be used by extending
the DFN to include additional physics such as degradation mechanisms, particle size distributions, non-spherical particles, etc. The
2+1D DFN model provides a framework into which such additional physical effects can be incorporated in a straightforward manner, providing  a computationally-efficient way of investigating how non-uniform cell use affects degradation, for example, and helping to rapidly assess new cell designs that aim to mitigate non-uniform aging of cells.

\vspace{2em}

\textbf{Acknowledgements.} This publication is based on work supported by the EPSRC Centre For Doctoral
Training in Industrially Focused Mathematical Modelling (EP/L015803/1) in collaboration with
Siemens Corporate Technology and BBOXX Ltd. The authors would also like to acknowledge funding
provided by The Faraday Institution, grant number EP/S003053/1, FIRG003.

\clearpage

\appendix

\section{Dimensionless Parameter Values and Variables}
In Table \ref{table:dimensionless_parameter_values} we provide the values of the dimensionless parameters, calculated from the dimensional parameters provided in Table~SM1. In Table \ref{table:dimensionless_variables}, we provide a glossary of the dimensionless variables, and their region of definition.

\begin{table}[htb]
	\centering
	\resizebox{\textwidth}{!}{%
	\begin{tabular}{c c p{5cm} c c c c c}
	\toprule
     Parameter & Expression & Interpretation & cn & n & s & p & cp  \\
    \midrule
    $L\ts{k}$ & $L\ts{k}^*/L_x^*$ & Ratio of region thickness to cell thickness & 0.1111 & 0.4444 & 0.1111 & 0.4444 & 0.1111 \\
    $A\ts{tab,k}$ & $A\ts{tab,k}^*/(L_x^*L^*)$ & Dimensionless tab area & - & 0.0324 & - & 0.0324 & - \\
    $\mathcal{C}\ts{k}$ & $\tau\ts{k}^*/\tau\ts{d}^*$ & Ratio of solid diffusion and discharge timescales &  - & 0.1135 $\mathcal{C}$ & - & 0.04425 $\mathcal{C}$ &- \\
    $\mathcal{C}\ts{r,k}$ & $\tau\ts{r,k}^*/\tau\ts{d}^*$ & Ratio of reaction and discharge timescales &  - & 0.0375 $\mathcal{C}$ & - & 1.5 $\mathcal{C}$ &- \\
   $\sigma\ts{k}$ & $(R^*T^*_{\infty}/F^*)/(I^*L^*_x/\sigma^*\ts{k})$ & Ratio of thermal voltage to the typical Ohmic drop in the solid & $2.84\times10^{8}/\mathcal{C}$ & $475.8/ \mathcal{C}$ & - & $47.58/ \mathcal{C}$ & $1.69\times10^{8}/\mathcal{C}$\\
    $a\ts{k}$ & $a\ts{k}^* R\ts{k}^*$ & Product of particle radius and surface area per unit volume   & - & 1.8 & - & 1.5 &-\\
   $\gamma\ts{k}$ & $c\ts{k,max}^*/c\ts{n,max}^*$ & Ratio of maximum lithium concentrations in solid & - & 1  & - & 2.0501&-\\
    $\rho\ts{k}$ & $\rho\ts{k}^*c\ts{p,k}^* / \rho\ts{eff}^*$ & Dimensionless volumetric heat capacity & 1.903 & 0.6403 & 0.1535 & 1.2605 & 1.3403 \\
    $\lambda\ts{k}$ & $\lambda\ts{k}^*/\lambda\ts{eff}^*$ & Dimensionless thermal conductivity & 6.7513 & 0.0286 & 0.0027 & 0.0354 & 3.9901\\
    \midrule
    $c\ts{k,0}$ & $c\ts{k,0}^*/c\ts{k,max}^*$ & Dimensionless initial lithium concentration in solid &-& 0.8 & - & 0.6 & -\\
    $T_0$ & $(T_0^* - T_\infty^*)/\Delta T^*$ & Dimensionless initial temperature & & & 0 & & \\
\midrule
    $\delta$ & $L_x^*/L^*$ & Aspect ratio  &&& &$1.642\times10^{-3}$ & \\
    $L_y$ & $L_y^*/L^*$ & Dimensionless cell width &&& &1.511 & \\
    $L_z$ & $L_z^*/L^*$ & Dimensionless cell height &&&& 1 & \\
    $\mathcal{C}\ts{e}$ & $\tau\ts{e}^*/\tau\ts{d}^*$ & \multicolumn{3}{l}{Ratio of electrolyte transport and discharge timescales   }&& $8.036\times10^{-3} \: \mathcal{C}$ & \\
    $\gamma\ts{e}$ & $c\ts{e,typ}^* / c\ts{n,max}^*$ & \multicolumn{3}{p{9cm}}{Ratio of maximum lithium concentration in the negative electrode solid and typical electrolyte concentration }&& $4.003 \times 10^{-2}$ & \\
    $\hat{\kappa}\ts{e}$ & $(R^*T^*_{\infty}/F^*)/(I^*L^*_x/\kappa^*\ts{e,typ})$ & \multicolumn{3}{p{9cm}}{Ratio of thermal voltage to the typical Ohmic drop in the electrolyte } & &  $4.981/\mathcal{C}$  & \\
    $\mathcal{B}$ & $I^*R^* T^*_{\infty}\tau\ts{th}^*/(\rho\ts{eff}^*F^*\Delta T^* L_x^*)$ & \multicolumn{3}{p{9cm}}{Dimensionless heat generation coefficient }&& $0.3608$ & \\
    $\mathcal{C}\ts{th}$ & $\tau\ts{th}^*/\tau\ts{d}^*$ & \multicolumn{3}{p{9cm}}{Ratio of planar thermal diffusion and discharge timescales}&& $0.02533 \mathcal{C}$ & \\
    $\Theta$ & $ \Delta T^* / T_{\infty}^*$ & \multicolumn{3}{p{9cm}}{Ratio of typical temperature variation and reference temperature}  && $0.008 \: \mathcal{C}$ & \\
    $h$ & $h^*L_x^*/\lambda\ts{eff}^*$ & \multicolumn{3}{p{9cm}}{Dimensionless heat transfer coefficient} && $3.7881\times10^{-5}$ & \\
    \bottomrule
    \end{tabular}}
        \caption{
         Typical dimensionless parameter values. Here $\mathcal{C}=I^*/(\SI{24}{\ampere \metre^{-2}})$ is the C-Rate where we have taken a 1C rate to correspond to a typical $x$-direction current density of $\SI{24}{\ampere \metre^{-2}}$. This is for a cell with an initial stoichiometry of $0.8$ in the negative electrode and $0.6$ in the positive electrode with a voltage cutoff of $\SI{3.2}{\volt}$.}
    \label{table:dimensionless_parameter_values}
\end{table}

\begin{table}[htb]
	\centering
	\begin{tabular}{c p{8cm} c}
	\toprule
     Symbol &  Interpretation & Region $\kin{$\cdots$}$  \\
    \midrule
    $\phi\ts{s,k}$ & Electric potential in solid & cn, n, p, cp \\
    $\phi\ts{e,k}$ & Electric potential in the electrolyte & n, s, p \\
    $\boldsymbol{i}\ts{s,k}$ & Current density in the solid & cn, n, p, cp \\
    $\boldsymbol{i}\ts{e,k}$ & Current density in the electrolyte & n, s, p \\
    $c\ts{s,k}$ & Lithium concentration in the active material & n, p \\
    $c\ts{e,k}$ & Lithium-ion concentration in the electrolyte & n, s, p \\
    $N\ts{s,k}$ & Lithium flux in the active material & n, p \\
    $\boldsymbol{N}\ts{e,k}$ & Lithium-ion flux in the electrolyte & n, s, p \\
    $j\ts{k}$ & Interfacial current density & n, p \\
    $j\ts{0,k}$ & Exchange current density & n, p \\
    $\eta\ts{k}$ & Surface reaction overpotential & n, p \\
    $U\ts{k}$ & Open circuit potential & n, p \\
    $T\ts{k}$ & Temperature & cn, n, s, p, cp \\
    $Q\ts{Ohm,k}$& Ohmic heating in solid & cn, cp \\
    $Q\ts{Ohm,k}$ & Combined Ohmic heating in solid and electrolyte & n, p \\
    $Q\ts{Ohm,k}$ & Ohmic heating in electrolyte & s \\
    $Q\ts{rxn,k}$ & Irreversible heating due to electrochemical reactions & n, p \\
    $Q\ts{rev,k}$ & Reverisble heating due to electrochemical reactions & n, p \\
    \bottomrule
    \end{tabular}
    \caption{Dimensionless variables}
    \label{table:dimensionless_variables}
\end{table}

\clearpage
\bibliographystyle{plain}
\bibliography{refs}

\end{document}


\maketitle


Here we provide a detailed exposition of the 3D DFN model expressed in terms of dimensional variables. The electrochemical model consists of a 3D version of the standard DFN model \cite{doyle1993}, with an additional equation relating to charge balance in the current collectors, which we model as Ohmic resistors. This is supplemented with an equation for energy balance, which includes source terms arising from Ohmic, irreverisble, and reversible heat generation with the cell. Descriptions of each of the cell components, along with any modelling assumptions, and the resulting governing equations,  are laid out in the following sections.

As indicated in Figure~1, the negative current collector, negative electrode, separator, positive electrode, and positive current collector are of thickness $L^*\ts{cn}$, $L^*\ts{n}$, $L\ts{s}^*$, $L\ts{p}^*$, and $L\ts{cp}^*$, respectively. We denote the distance between the negative and positive current collectors by $L_x^*=L\ts{n}^*+L\ts{s}^*+L\ts{p}^*$, the width of the cell by $L_y^*$ and the height of the cell by $L_z^*$. The active material particles in the negative and positive electrodes are assumed to be spheres with radii $R\ts{n}^*$ and $R\ts{p}^*$, respectively. We use the spatial coordinates $\boldsymbol{x}^*=(x^*, y^*, z^*)\in[-L^*\ts{cn},L^*_x+L^*\ts{cp}]\times[0, L^*_y]\times[0, L^*_z]$ to indicate the location through the thickness, width, and height of the cell, and the spatial coordinate $r^*\in[0,R\ts{k}^*]$, $\kin{n, p}$, to indicate the location within the particles of active material. Time is denoted by $t^*$. We define 
\begin{align*}
    \Omega^*\ts{cn} &=  [-L\ts{cn}^*,0]\times\Omega^*, &
    \Omega^*\ts{n} &=  [0,L\ts{n}^*]\times\Omega^*, &
    \Omega^*\ts{s} &=  [L\ts{n}^*,L_x^*-L\ts{p}^*]\times\Omega^*, \\
    \Omega^*\ts{p} &=  [L_x^*-L\ts{p}^*,L_x^*]\times\Omega^*, &
    \Omega^*\ts{cp} &=  [L_x^*, L_x^*+L\ts{cp}^*]\times\Omega^*,
\end{align*}
corresponding to the negative current collector, negative electrode, separator, positive electrode, and positive current collector, respectively, where $\Omega^* =  [0,L_y^*]\times[0,L_z^*]$ is the projection of the battery onto the $(y^*,z^*)$-plane. 
Similar to the main document, we  use the notation $\partial\Omega^*\ts{tab,k}$ to refer to the negative and positive tabs ($\kin{cn, cp}$), $\partial\Omega^*\ts{ext,k}$ to refer to the external boundaries of region $\kin{cn, n, s, p, cp}$, and $\partial\Omega^*_{\text{k}_1, \text{k}_2}$ to refer to the interface between regions $\text{k}_1$ and $\text{k}_2$.

\section{Current collectors}
In the current collectors conservation of charge along with Ohm's law gives
\begin{subequations}\label{eqn:dim:current_collectors}
\begin{align}
    \nabla^* \cdot \boldsymbol{i}^*\ts{s,ck} &= 0,  && \kin{n, p}, \\
    \boldsymbol{i}^*\ts{s,ck} &= - \sigma^*\ts{ck} \nabla^* \phi^*\ts{s,ck}, &&  \kin{n, p},
\end{align}
where $\boldsymbol{i}^*\ts{s,ck}$ is the current density, $\phi^*\ts{s,ck}$ is the electrical potential, and $\sigma^*\ts{ck}$ is the electrical conductivity.
We assume no charge is exchanged through the external boundary of the current collectors except at the tabs so that
\begin{align}
    \boldsymbol{i}^*\ts{s,ck}\cdot\boldsymbol{n} &= 0, \quad &&\boldsymbol{x}^* \in \partial \Omega^*\ts{ext,ck}, \quad &&&\kin{n, p}.
\end{align}
We need to set a reference potential, or ground state, against which other potentials are measured. We assume a uniform potential across the surface of the negative tab and set this potential to be zero
\begin{equation}
    \phi^*\ts{s,cn} = 0, \quad \boldsymbol{x}^* \in \partial \Omega^*\ts{tab,n}.
\end{equation}
The boundary condition that describes the charging rate of the cell can be implemented in several ways. If the voltage is specified then we can take the potential at the positive tab to be given. Alternatively, the applied current may be specified, which can be implemented in different ways: either by assuming the electric potential at the positive tab is spatially uniform, but unknown, and letting the total applied current be given; or by assuming that the applied current is uniform across one tab and letting the voltage be determined by the average potential on the tab. Adopting the former approach provides the following two conditions describing the current flow through the tabs
\begin{equation}
    \label{bc:tab_integral}
      \int_{\partial\Omega\ts{tab,n}} \boldsymbol{i}^*\ts{s,cn}\cdot\boldsymbol{n} \dd{x} = -  I^*\ts{app}, \quad \text{and hence} \quad
      \int_{\partial\Omega\ts{tab,p}} \boldsymbol{i}^*\ts{s,cp}\cdot\boldsymbol{n} \dd{x} = I^*\ts{app},
\end{equation}
where $\boldsymbol{n}$ is the outward pointing unit normal. In practice the first of these conditions is not needed, since we set the potential to zero on the negative tab, and conservation of current ensures the integral conditions \eqref{bc:tab_integral} are satisfied. If instead we assume a uniform current density across the tabs the integral condition at the positive tab is replaced with
\begin{equation}
    \boldsymbol{i}^*\ts{s,cp}\cdot\boldsymbol{n} = -\frac{I^*\ts{app}}{A^*\ts{tab,p}}, \quad \boldsymbol{x}^* \in \partial \Omega^*\ts{tab,p},
\end{equation}
where $A\ts{tab,k}^* = L^*\ts{ck}L^*\ts{tab,k}$ is the area of tab k, $L^*\ts{ck}$ is the current collector thickness, and $L^*\ts{tab,k}$ is the tab width. Finally, the electrical potential and the normal component of the current density are continuous across the current collector--electrode interface
\begin{align}
    \phi^*\ts{s,ck} &= \phi^*\ts{s,k}, \quad &&\boldsymbol{x}^* \in \partial \Omega^*\ts{ck,k}, \quad &&& \kin{n, p}, \label{eqn:cc-e:boundary_a} \\
    \boldsymbol{i}^*\ts{s,ck}\cdot\boldsymbol{n}
    &= \boldsymbol{i}^*\ts{s,k}\cdot\boldsymbol{n}, \quad &&\boldsymbol{x}^* \in \partial \Omega^*\ts{ck,k}, \quad &&& \kin{n, p}, \label{eqn:cc-e:boundary_b}
\end{align}
where $\phi\ts{s,k}$ and $\boldsymbol{i}\ts{s,k}$ are the electrical potential and current density in electrode k.
\end{subequations}

\section{Electrodes}
The electrodes in a lithium-ion cell are electrically conducting porous materials which consist of active material, where lithium can intercalate, and an electrically conducting binder material. The pores of the electrode are flooded with an electrolyte which will be described in detail in Section~\ref{sec:electrolyte}. Within the framework of the DFN model, the active material is treated as spherical particles, with diffusion being the transport mechanism for lithium transport within the solid \cite{doyle1993}. The behaviour within the particle is assumed to be spherically symmetric. The timescale associated with diffusion within the active material is long compared with the timescale associated with the transport of electrons, and the model therefore requires two components which describe processes occurring over disparate lengthscales: a macroscale description of charge conservation (electron transport), and a microscale description of mass conservation (lithium transport) \cite{doyle1993}.

\subsection{Charge conservation in the electrodes}
The current in the electrodes is described by Ohm's law. To account for the transfer of current between the electrode and electrolyte (which occurs via electrochemical reactions, which are discussed shortly), an additional current source/sink term is included. The governing equations for current in the electrode are therefore
\begin{subequations}\label{eqn:dim:electrode_current}
\begin{align}
    \nabla^* \cdot \boldsymbol{i}^*\ts{s,k} &= -a^*\ts{k} j^*\ts{k},  && \kin{n, p}, \\
    \boldsymbol{i}^*\ts{s,k} &= - \sigma^*\ts{s,k} \nabla^* \phi^*\ts{k}, &&  \kin{n, p},
\end{align}
where $a^*\ts{k}$ is the surface area per unit volume  of the electrode, and $j\ts{k}^*$ is the interfacial current density. On the boundaries between the electrodes and current collectors the continuity conditions \eqref{eqn:cc-e:boundary_a} and \eqref{eqn:cc-e:boundary_b} hold. On the external boundaries of the cell no charge is transferred
\begin{align}
    \boldsymbol{i}^*\ts{s,k}\cdot\boldsymbol{n} &= 0, \quad &&\boldsymbol{x}^* \in \partial \Omega^*\ts{ext,k}, \quad &&&\kin{n, p}.
\end{align}
Finally, the separator is taken to be electrically insulating so that no charge is transferred from the electrodes to the separator (charge can be transferred through the separator region but only by ion transport in the electrolyte which floods the pores of the separator). Therefore, on the electrode--separator boundary, we have
\begin{align}
    \boldsymbol{i}^*\ts{s,k}\cdot\boldsymbol{n} &= 0, \quad &&\boldsymbol{x}^* \in \partial \Omega^*\ts{k,s}, \quad &&& \kin{n, p}.
\end{align}
\end{subequations}

\subsection{Mass conservation in the active material}
As in \cite{doyle1993}, we treat the active material on the microscale as spherical particles of uniform radius, in which spherically symmetric diffusion of lithium is described by Fick's law
\begin{subequations}\label{eqn:dim:active_material}
\begin{align}
     \pdv{c^*\ts{s,k}}{t^*} &= -\frac{1}{(r^*)^2}\pdv{r^*}\bigg((r^*)^2 N\ts{s,k}^*\bigg), \quad N\ts{s,k}^* = -D^*\ts{s,k}(c^*\ts{s,k}, T^*\ts{k})\pdv{c^*\ts{s,k}}{r^*}, \quad && \kin{n, p},
\end{align}
where $c\ts{s,k}^*$ is the concentration of lithium in the active material, $N\ts{s,k}^*$ is the flux of lithium ions in the active material, $D\ts{s,k}^*(c^*\ts{s,k}, T^*\ts{k})$ is the diffusivity of lithium in the active material, $T^*\ts{k}$ is the (macroscopic) temperature, $r^*$ is the radial spatial coordinate, and $t^*$ is time. We assume that the particle is entirely surrounded by electrolyte and that lithium transfer with the electrolyte occurs uniformly across each particle's outer surface, giving
\begin{align}
    N\ts{s,k}^*\big|_{r^*=0} = 0, \quad N\ts{s.k}^*\big|_{r^*=R^*\ts{k}} = \frac{j^*\ts{k}}{F^*}.  \quad \kin{n, p},
\end{align}
where $F^*$ is Faraday's constant. Further, we assume that the concentration within the particles in each electrode is initially uniform in space
\begin{equation}
    c^*\ts{s,k}\big|_{t^*=0} = c^*\ts{s,k,$0$},
\end{equation}
where $c^*\ts{s,k,$0$}$ is a constant. It should be noted that this microscale model for the active material holds at every point $\boldsymbol{x}^*\in\Omega\ts{k}^*$ for $\kin{n, p}$ of the macroscale model. In this sense, the radial direction $r^*$ can be viewed as a `pseudodimension'. This also means that $c\ts{s,k}^*$ is in general a function of $r^*$, $\boldsymbol{x}^*$, and $t^*$.
\end{subequations}

\section{Electrolyte}\label{sec:electrolyte}
The electrolyte is taken to be a moderately concentrated binary electrolyte consisting of lithium-ions, a single anion species, and a solvent species. We refer to the description of the electrolyte as moderately concentrated in the sense that species interactions are accounted for though Stefan--Maxwell constitutive laws for the species fluxes, but that the total concentration of the electrolyte is approximately that of the solvent so that the electrolyte density can be considered constant. This description provides a migration--diffusion equation for both of the ion species. After appealing to charge neutrality, linear combinations of these two equations yield an equation for the current in the electrolyte as well as an effective diffusion equation for the concentration of both anions and lithium-ions.

\subsection{Charge conservation in the electrolyte}
In the electrolyte current is driven by concentration gradients and the effects of interacting species must be accounted for, resulting in a modified Ohm's Law. Charge transfer with the electrode solid material, through electrochemical reactions, is accounted for by an additional source term. In the separator, this source term is zero since no charge transfer occurs between the electrolyte and the separator material. The governing equations for the current in the electrolyte are
\begin{subequations}\label{eqn:dim:electrolyte_current}
\begin{align}
    \label{eqn:dim:electrolyte:current:electrodes}
    \nabla^* \cdot \boldsymbol{i}^*\ts{e,k} &= a^*\ts{k} j^*\ts{k},\qquad \qquad   \kin{n, p}, \\
    \label{eqn:dim:electrolyte:current:sep}
    \nabla^* \cdot \boldsymbol{i}^*\ts{e,s} &= 0,  \\
    \boldsymbol{i}^*\ts{e,k} &= \epsilon^{\text{b}}\ts{k} \kappa\ts{e}^*(c^*\ts{e,k}, T^*\ts{k}) \left( - \nabla^* \phi^*\ts{e,k} + 2(1-t^+)\frac{R^*T^*\ts{k}}{F^*} \nabla^*\left(\log(c^*\ts{e,k})\right) \right),\\
    &\nonumber \hspace{5cm} \kin{n, s, p},
\end{align}
where $\boldsymbol{i}\ts{e,k}^*$ is the current in the electrolyte, $\epsilon\ts{k}$ is the electrolyte volume fraction, $b$ is the Bruggeman coefficient, $c^*\ts{e,k}$ is the lithium-ion concentration, $\kappa^*\ts{e}(c^*\ts{e,k}, T^*\ts{k})$ is the electrolyte conductivity, $\phi^*\ts{e,k}$ is the electrical potential in the electrolyte, $t^+$ is the transference number, and $R^*$ is the universal gas constant. No charge is transferred directly from the electrolyte into the current collectors, so that
\begin{align}
    \boldsymbol{i}^*\ts{e,k}\cdot\boldsymbol{n} &= 0, \quad &&\boldsymbol{x}^* \in \partial \Omega^*\ts{ck,k}, \quad &&& \kin{n, p},
\end{align}
and no charge is transferred from the electrolyte through the external boundaries
\begin{align}
    \boldsymbol{i}^*\ts{e,k}\cdot\boldsymbol{n} &= 0, \quad &&\boldsymbol{x}^* \in \partial \Omega^*\ts{ext,k}, \quad &&&\kin{n, s, p}.
\end{align}
Finally, on the electrode--separator boundaries the the electrical potential and current in the electrolyte must be continuous
\begin{align}
    \phi^*\ts{e,k} &= \phi^*\ts{e,s}, \quad &&\boldsymbol{x}^* \in \partial \Omega^*\ts{k,s}, \quad &&& \kin{n, p}, \\
    \boldsymbol{i}^*\ts{e,k}\cdot\boldsymbol{n} &= \boldsymbol{i}^*\ts{e,s}\cdot\boldsymbol{n}, \quad &&\boldsymbol{x}^* \in \partial \Omega^*\ts{k,s}, \quad &&& \kin{n, p}.
\end{align}
Note that the electrical potential in the electrolyte is only determined up to a constant, which is is determined relative to  the reference electrode potential on the negative tab through the electrochemical reactions.
\end{subequations}

\subsection{Mass conservation in the electrolyte}
The concentration of lithium ions in the electrolyte is determined by solving an effective diffusion equation with an additional source term describing lithium transfer to the active material
\begin{subequations}\label{eqn:dim:electrolyte_diffusion}
\begin{align}
     \epsilon\ts{k} \pdv{c^*\ts{e,k}}{t^*} &= -\nabla^* \cdot \boldsymbol{N}^*\ts{e,k} + \frac{1}{F^*} \nabla^*\cdot\boldsymbol{i}\ts{e,k}^*, \quad && \kin{n, s, p},\\
      \boldsymbol{N}^*\ts{e,k} &= -\epsilon^{\text{b}}\ts{k} D^*\ts{e}(c^*\ts{e,k}, T^*\ts{k})\nabla^* c^*\ts{e,k} + \frac{t^+}{F^*} \boldsymbol{i}\ts{e,k}^*, \quad && \kin{n, s, p}
\end{align}
where we have used \eqref{eqn:dim:electrolyte:current:electrodes} and \eqref{eqn:dim:electrolyte:current:sep} to give a concise description of the flux of lithium ions across the interface between the electrolyte and solid particle. Here, $\boldsymbol{N}^*\ts{e,k}$ is the lithium-ion flux in the electrolyte and $D\ts{e}^*(c^*\ts{e,k}, T^*\ts{k})$ is the diffusivity of the electrolyte. There is no flux of lithium ions from the electrolyte into the current collectors or out of the external boundaries
\begin{align}
    \boldsymbol{N}^*\ts{e,k}\cdot\boldsymbol{n} &= 0, \quad &&\boldsymbol{x}^* \in \partial \Omega^*\ts{ck,k}, \quad &&& \kin{n, p}, \\
    \boldsymbol{N}^*\ts{e,k}\cdot\boldsymbol{n} &= 0, \quad &&\boldsymbol{x}^* \in \partial \Omega^*\ts{ext,k}, \quad &&& \kin{n, s, p}.
\end{align}
 We also require that the concentration and normal flux of lithium ions to be continuous across the electrode/separator boundaries
\begin{align}
    c^*\ts{e,k} &= c^*\ts{e,s} \quad &&\boldsymbol{x}^* \in \partial \Omega^*\ts{k,s}, \quad &&& \kin{n, p}, \\
    \boldsymbol{N}^*\ts{e,k}\cdot\boldsymbol{n} &= \boldsymbol{N }^*\ts{e,s}\cdot\boldsymbol{n}, \quad &&\boldsymbol{x}^* \in \partial \Omega^*\ts{k,s}, \quad &&& \kin{n, p}, \\
\end{align}
and assume that the concentration of lithium ions in the electrolyte is initially uniform in space
\begin{align}
    c^*\ts{e,k}\big|_{t^*=0} = c^*\ts{e,$0$}, \quad \kin{n, s, p},
\end{align}
where $c^*\ts{e,$0$}$ is constant.
\end{subequations}

\section{Electrochemical reactions}
The electrochemical reactions are modelled using symmetric Butler--Volmer kinetics. The reaction flux density, $j^*\ts{k}$, is then given by
\begin{subequations}\label{eqn:dim:electrochemical_reactions}
\begin{align}
    &j^*\ts{k} =  j^*\ts{$0$,k}\sinh\left(\frac{F^*\eta^*\ts{k}}{2R_g^*T^*\ts{k}} \right),
     && \kin{n,p}, \\
    &j^*\ts{$0,$k} = m^*\ts{k}(T^*\ts{k}) (c^*\ts{s,k})^{1/2} (c^*\ts{s,k,max} - c^*\ts{s,k})^{1/2}(c^*\ts{e,k})^{1/2} \bigg|_{r^*=R^*\ts{k}}  && \kin{n, p}, \\
    &\eta^*\ts{k} = \phi^*\ts{s,k} - \phi^*\ts{e,k} - U^*\ts{k}\big|_{r^*=R^*\ts{k}},  && \kin{n, p},
\end{align}
\end{subequations}
where $j^*\ts{k}$ is the interfacial current density, $\eta^*\ts{k}$ is the surface reaction overpotential, $U^*\ts{k}$ is the open circuit potential, and $m^*\ts{k}$ is the kinetic rate constant.

\section{Energy conservation}
The governing equation for energy conservation is
\begin{subequations}\label{eqn:dim:thermal}
\begin{equation}
    \rho^*\ts{k} c^*\ts{p,k} \pdv{T^*\ts{k}}{t^*} = \nabla^*\cdot\left(\lambda^*\ts{k} \nabla^*T^*\ts{k}\right) +  Q^*\ts{Ohm,k} + Q^*\ts{rxn,k} + Q^*\ts{rev,k}, \quad \kin{cn, n, s, p, cp},
\end{equation}
where $\rho^*\ts{k}$ is the density, $c^*\ts{p,k}$ is the specific heat, $\lambda^*\ts{k}$ is the thermal conductivity. Within the electrode region, the model accounts for Ohmic heating $Q^*\ts{Ohm,k}$ due to resistance in the solid and electrolyte, irreverisble heating due to electrochemical reactions $Q^*\ts{rxn,k}$, and reversible heating due to entropic changes in the the electrode $Q^*\ts{rev,k}$ \cite{bernardi1985}, given by
\begin{align}
    Q^*\ts{Ohm,k} &= - \left(\boldsymbol{i}^*\ts{s,k} \cdot\nabla^*\phi\ts{s,k} + \boldsymbol{i}^*\ts{e,k} \cdot\nabla^* \phi^*\ts{e,k}\right), && \kin{n, p},\\
    Q^*\ts{rxn,k} &= a^*\ts{k} j^*\ts{k} \eta^*\ts{k}, && \kin{n,p}, \\
    Q^*\ts{rev,k} &= a^*\ts{k} j^*\ts{k} T^*\ts{k} \pdv{U^*\ts{k}}{T^*\ts{k}}\bigg|_{T^*\ts{k}=T_\infty^*}, && \kin{n, p}.
\end{align}
However, in the current collectors and separator there is no heat generation due to electrochemical effects, and we need only consider the Ohmic heat generation terms given by
\begin{equation}
    Q^*\ts{Ohm,k} = -\boldsymbol{i}^*\ts{s,k}\cdot\nabla^*\phi\ts{s,k}, \quad \kin{cn, cp},\quad \quad Q^*\ts{Ohm,s} = -\boldsymbol{i}^*\ts{e,s}\cdot\nabla^*\phi\ts{e,s}.
\end{equation}

For the thermal part of the problem we assume Newton cooling on all boundaries, including the tabs,
\begin{equation}
    \label{bc:thermal_ext}
    -\lambda^*\ts{k}\nabla^* T^*\ts{k} \cdot \boldsymbol{n} = h^*(T^*\ts{k} - T^*_{\infty}), \quad \boldsymbol{x}^* \in \partial \Omega^*\ts{ext}, \quad  \kin{cn, n, s, p, cp},
\end{equation}
where $h^*$ is the (possibly spatially dependent) heat transfer coefficient.

We require the  temperature and heat flux to be continuous at the interfaces between the components of the cell
\begin{align}
    T^*\ts{ck} = T^*\ts{k}, \quad \lambda^*\ts{ck}\nabla^* T^*\ts{ck} \cdot\boldsymbol{n} &= \lambda^*\ts{k}\nabla^* T^*\ts{k}\cdot\boldsymbol{n} \quad &&\boldsymbol{x}^* \in \partial \Omega^*\ts{ck,k}, \quad &&& \kin{n, p}, \\
    T^*\ts{k} = T^*\ts{s}, \quad \lambda^*\ts{k}\nabla^* T^*\ts{k} \cdot\boldsymbol{n} &= \lambda^*\ts{s}\nabla^* T^*\ts{s}\cdot\boldsymbol{n} \quad &&\boldsymbol{x}^* \in \partial \Omega^*\ts{k,s}, \quad &&& \kin{n, p},
\end{align}
and prescribe an initial uniform temperature $T^*\ts{0}$, that is
\begin{equation}
    T^*\ts{k}\big\vert_{(t^*=0)} = T^*\ts{0}.
\end{equation}
\end{subequations}


\section{Non-dimensionlisation}
In order to facilitate the asymptotic analysis in the main paper we write the model in dimensionless form. This is achieved by introducing characteristic scales for each of the variables, in a straightforward extension of \cite{Marquis2019}. For reference, all of the parameters and scalings are defined in Table~\ref{table:dimensional_parameter_values}, along with typical values for a carbon negative current collector, graphite negative electrode, LiPF$_6$ in EC:DMC electrolyte, lithium cobalt oxide positive electrode, and aluminium positive current collector, adapted from \cite{Moura2017, SMouraGithub, DUALFOIL}. The through-cell $x$ coordinate is scaled with $L_x^*$ and plane $y$, $z$ coordinates are scaled with a typical transverse dimension $L^*$ (which could be $L^*_y$, $L^*_z$ or $(L^*_y L^*_z)^{1/2}$ for example). In each of the particles we scale lengths by the particle radius. The tab surface areas are naturally scaled as $A^*\ts{tab,k} = L^*_xL^* A\ts{tab,k}$. As our typical timescale we take a representative  discharge time of the cell $\tau^*\ts{d}$, which is defined in Table~\ref{table:timescales}. The current densities in both the electrode and electrolyte are scaled by a typical operating current density $I^*$. Naturally, the applied current scales as $I\ts{app}^* = I^*(L^*)^2 I\ts{app}$. We choose to scale the potentials, overpotentials, and open-circuit potentials (OCP) by a typical potential scale $\Phi^*$, which we take to be the thermal voltage $\Phi^* = R^*T^*_{\infty}/F^*$. 
The scaled potential in the negative electrode is measured relative to ground (\SI{0}{V}), whereas the scaled potential in the positive electrode is measured relative to the reference open-circuit voltage (OCV) $U^*\ts{ref} = U^*\ts{p,ref} - U^*\ts{n,ref}$, which is the voltage measured across the cell when no current is applied. The scaled potential in the electrolyte is measured relative to the reference OCP in the negative electrode $U^*\ts{n,ref}$. The temperature is measured relative to the ambient temperature $T^*_\infty$, and we scale the deviation by some typical temperature difference $\Delta T^*$,  to be determined later. The lithium concentration in the electrolyte is scaled by the typical value $c\ts{e,typ}^*$, whereas the lithium concentrations in the solid particles are scaled with their maximum concentrations, $c^*\ts{s,k,max}$. Finally, the electrolyte conductivity, electrolyte diffusivity, solid diffusivity, and reaction rates are all scaled by their typical values $\kappa^*\ts{e,typ}$, $D^*\ts{e,typ}$, $D\ts{s,k,typ}^*$, and $m\ts{k,typ}^*$, respectively. Fluxes then scale in the natural way. To summarise, the dimensionless variables are related to their dimensional counterparts in the following way:
\[
        \begin{aligned}
        &\text{global} &&x^* = L^*_x x, \quad
        y^* = L^* y, \quad
        z^* = L^* z, \quad
        t^* = \tau^*\ts{d} t, \\
        & &&I\ts{app}^* = I^* (L^*)^2 I\ts{app} \quad
        \kappa^*\ts{e} = \kappa\ts{typ}^*\kappa\ts{e}, \quad
        D^*\ts{e} = D^*\ts{e,typ}D\ts{e};
        \\[10pt]
        &\text{for } \kin{n, p}: \
        &&r^*\ts{k} = R^*\ts{k} r\ts{k}, \quad
        c\ts{s,k}^* = c^*\ts{s,k,max}c\ts{s,k} \quad D\ts{s,k}^* = D\ts{s,k,typ}^* D\ts{s,k}, \\
        & && \qquad N\ts{s,k}^* = \frac{D\ts{s,k,typ}^*c\ts{s,k,max}^*}{R\ts{k}^*}N\ts{s,k}, \\
        & && j^*\ts{k} =  \frac{I^*}{a^*\ts{k}L^*_x}j\ts{k}, \quad j^*\ts{$0,$k} = \frac{I^*}{a^*\ts{k}L^*_x}j\ts{$0,$k}, \quad m\ts{k}^* = m\ts{k,typ}^*m\ts{k}, \\
        & && \eta^*\ts{k} = \frac{R^* T^*_{\infty}}{F^*}\eta\ts{k}, \quad
        U^*\ts{k} = U^*\ts{k,ref} + \frac{R^* T^*_{\infty}}{F^*}U\ts{k}; \quad
         \\[10pt]
        &\text{for } \kin{cn, n, p, cp}: \
        && \boldsymbol{i}^*\ts{s,k} = I^* \boldsymbol{i}\ts{s,k}; \\[10pt]
        &\text{for } \kin{cn, n}: \
        && \phi^*\ts{s,k} = \frac{R^* T^*_{\infty}}{F^*}\phi\ts{s,k}; \\[10pt]
        &\text{for } \kin{p, cp}: \
        && \phi^*\ts{s,k} = \left(U^*\ts{p,ref} - U^*\ts{n,ref}\right) + \frac{R^* T^*_{\infty}}{F^*}\phi\ts{s,k}; \\[10pt]
        &\text{for } \kin{n, s, p}: \
        && c^*\ts{e,k} = c^*\ts{e,typ}c\ts{e,k}, \quad
        \boldsymbol{N}^*\ts{e,k} = \frac{D\ts{e,typ}^*c\ts{e,typ}^*}{L_x^*}\boldsymbol{N}\ts{e,k}, \\
        & && \phi^*\ts{e,k} = - U^*\ts{n,ref} + \frac{R^* T^*}{F^*}\phi\ts{e,k}, \quad
        \boldsymbol{i}^*\ts{e,k} = I^* \boldsymbol{i}\ts{e,k}; \\[10pt]
        &\text{for } \kin{cn, n, s, p, cp}: \
        && T^*\ts{k} = (\Delta T^*) T\ts{k} + T^*_\infty.
    \end{aligned}
\]

Through non-dimensionalisation, we identify a number of key timescales in our model which relate to various physical processes. These are provided and interpreted in Table~\ref{table:timescales}. In addition, we identify a number of dimensionless parameters which are presented alongside their respective interpretations and typical values in Table~2. The non-dimensional groupings and relevant timescales, with the exception of those relating to thermal effects, are the same as those found in \cite{Marquis2019} so are not discussed here.

So far, we have left the typical temperature rise $\Delta T^*$ undetermined. Choosing to balance the heat generation terms with the time derivative of the temperature gives unreasonably high values of $\Delta T^*$, as found in \cite{baker1999}. If we instead choose to balance the heat generation term with the surface cooling term, i.e. assume the dimensionless parameter $\mathcal{B} \sim \hp$, we find that $\Delta T^* \sim I^*\Phi^*/h^* \approx \SI{1}{K}$, which is in line with the typical local variation in temperature observed in experiments at moderate discharge rates of 1C or less \cite{kosch2018, GOUTAM2017, kumaresan2008thermal}. Thus the temperature is approximately quasi-static in this regime.


\section{Parameter values}
Parameters values are taken from \cite{SMouraGithub} and given in Table~\ref{table:dimensional_parameter_values}. The parameters are for a carbon negative current collector, graphite negative electrode, LiPF$_6$ in EC:DMC electrolyte, LCO positive electrode, and aluminium positive current collector.
The functional form of any parameters that depend on concentration is same as that found in \cite{SMouraGithub}. Any parameters that depend on temperature have an Arrhenius dependence, i.e. for a parameter $\psi^*$ that depends on concentration $c^*$ and temperature $T^*$ we have
\begin{equation*}
    \psi^*(c^*, T^*) = \tilde{\psi}(c^*)\exp\left(\frac{E^*}{R^*} \left( \frac{1}{T^*_{\infty}} - \frac{1}{T^*}\right)\right),
\end{equation*}
where $E^*$ is the activation energy and $R^*$ is the universal gas constant.
The effective volumetric heat capacity and thermal conductivity given in Table~\ref{table:dimensional_parameter_values} are defined as
\begin{equation}
    \rho^*\ts{eff} = \frac{\sum\limits\ts{k} \rho^*\ts{k} c^*\ts{p,k} L^*\ts{k}}{\sum\limits\ts{k} L^*\ts{k}} \quad \text{and} \quad \lambda^*\ts{eff} = \frac{\sum\limits\ts{k} \lambda^*\ts{k} L^*\ts{k}}{\sum\limits\ts{k} L^*\ts{k}},
\end{equation}
respectively.

The corresponding dimensionless parameters are given in Table~2, with the different (dimensional) timescales listed in Table~\ref{table:timescales}.

\begin{table}[htb]
	\centering
	\resizebox{\textwidth}{!}{%
	\begin{tabular}{c c p{5cm} c c c c c}
	\toprule
     Parameter & Units & Description & cn & n & s & p & cp \\
    \midrule
     $L^*\ts{k}$ & $\SI{}{\micro\metre}$ & Region thickness & 25 & 100 & 25 & 100 & 25 \\
     $L^*\ts{tab,ck}$ & $\SI{}{\milli\metre}$ & Tab width & - & 40 & - & 40 & - \\
    $c^*\ts{e,typ}$ & $\SI{}{\mol.\metre^{-3}}$ & Typical lithium concentration in electrolyte & - & $1\times10^3$ & $1\times10^3$ & $1\times10^3$ & - \\
    $D\ts{e,typ}^*$ & $\SI{}{\metre^{2}.\second^{-1}}$ & Typical electrolyte diffusivity  &- & $5.34\times10^{-10}$ & $5.34\times10^{-10}$ & $5.34\times10^{-10}$ & - \\
    $\epsilon\ts{k}$ &-  & Electrolyte volume fraction & - & $0.3$ & 1 & $0.3$ & - \\
    $c^*\ts{s,k,max}$ & $\SI{}{\mol.\metre^{-3}}$ & Maximum lithium concentration in solid  & - & $2.498\times 10^4$ & - & $5.122\times 10^4$ & - \\
     $\sigma^*\ts{k}$ & $\SI{}{\ohm^{-1}.\metre^{-1}}$ & Solid conductivity  & $5.96\times10^{7}$ & 100 & - & 10 & $3.55\times10^7$ \\
     $D^*\ts{s,k,typ}$ & $\SI{}{\metre^{2}.\second^{-1}}$ & Typical solid diffusivity  & - &  $3.9\times10^{-14}$ & - &  $1\times10^{-13}$ & - \\
     $R^*\ts{k}$ & $\SI{}{\micro\metre}$ & Particle radius  & - & 10 & - & 10 & -  \\
     $a^*\ts{k}$ & $\SI{}{\micro\metre^{-1}}$ & Electrode surface area per unit volume  & - & 0.18 & - & 0.15 & - \\
     $m^*\ts{k,typ}$ & $\SI{}{\ampere.\metre^{-2}.(\metre^3.\mol^{-1})^{1.5}}$ & Typical reaction rate  & - & $2\times10^{-5}$ & - & $6\times10^{-7}$ & - \\
     $\rho^*\ts{k}$ & $\SI{}{\kilogram.\metre^{-3}}$ & Density  & 8954 & 1657 & 397 & 3262 & 2707 \\
     $c^*\ts{p,k}$ & $\SI{}{\joule.\kilogram^{-1}.\kelvin^{-1}}$ & Specific heat capacity  & 385 & 700 & 700 & 700 & 897 \\
     $\lambda^*\ts{k}$ & $\SI{}{\watt.\metre^{-1}.\kelvin^{-1}}$ & Thermal conductivity  & 401 & 1.7 & 0.16 & 2.1 & 237 \\
   \midrule
   $E^*_{m\ts{k}^*}$ & $\SI{}{\joule.\mol^{-1}}$ & Activation energy for reaction rate & - & $3.748 \times 10^4$ & - & $3.957 \times 10^4$ & - \\
    $E^*_{D\ts{e}^*}$ & $\SI{}{\joule.\mol^{-1}}$ & Activation energy for electrolyte diffusivity &  &  & $3.704 \times 10^4$ &  &  \\
    $E^*_{\kappa\ts{e}^*}$ & $\SI{}{\joule.\mol^{-1}}$ & Activation energy for electrolyte conductivity &  &  & $3.470 \times 10^4$ &  &  \\
    \midrule
    $c^*_{k,0}$ & $\SI{}{\mol.\metre^{-3}}$ & Initial lithium concentration in solid  & - & $1.999\times10^4$ & - & $3.073\times10^4$ & - \\
    $T^*_0$ & $\SI{}{\kelvin}$ & Initial temperature & & & 298.15 & & \\
      \midrule
     $F^*$ & $\SI{}{\coulomb.\mol^{-1}}$ & Faraday's constant  & & & 96487 & & \\
     $R_g^*$ & $\SI{}{\joule.\mol^{-1}.\kelvin^{-1}}$ & Universal gas constant  &&& 8.314 && \\
     $T^*_\infty$  & $\SI{}{\kelvin}$ & Reference temperature &&& 298.15 && \\
     $b$ & - & Bruggeman coefficient  &&& 1.5 &&  \\
     $t^+$ & - & Transference number &&& 0.4 && \\
     $L^*_x$ & $\SI{}{\micro\metre}$ & Cell thickness  &&& 225 && \\
     $L^*_y$ & $\SI{}{\milli\metre}$ & Cell width  &&& 207 && \\
     $L^*_z$ & $\SI{}{\milli\metre}$ & Cell height  &&& 137 && \\
    $I^*$ & $\SI{}{\ampere.\metre^{-2}}$ & Typical current density  &&&  24 && \\
    $h^*$ & $\SI{}{\watt.\metre^{-2}.\kelvin^{-1}}$ & Heat transfer coefficient  &&& 10 && \\
    $\Delta T^*$ & $\SI{}{\kelvin}$ & Typical temperature variation  &&& 2.4 && \\
    $\rho\ts{eff}^*$ & $\SI{}{\joule.\kelvin^{-1}.\metre^{-3}}$ & \multicolumn{2}{l}{Effective volumetric heat capacity} & & $1.812 \times 10^{6}$  & & \\
   $\lambda\ts{eff}^*$ & $\SI{}{\watt.\metre^{1-}.\kelvin^{-1}}$ & Effective thermal conductivity & & & 59.396  & & \\
    \bottomrule
	\end{tabular}}
    \caption{Typical dimensional parameter values taken from \cite{SMouraGithub}. The parameters are for a carbon negative current collector, graphite negative electrode, LiPF$_6$ in EC:DMC electrolyte, LCO positive electrode, and aluminium positive current collector. } \label{table:dimensional_parameter_values}
\end{table}

\begin{table}[htb]
	\centering
	\resizebox{\textwidth}{!}{%
	\begin{tabular}{c c p{7cm} c}
	\toprule
     Symbol & Expression & Interpretation & Value (s) \\
    \midrule
    $\tau\ts{d}^*$ & ${F^* c^*\ts{n,max} L^*_x}/{I^*}$ & Discharge timescale &  $2.260 \times 10^4 / \mathcal{C}$ \\
    $\tau\ts{n}^*$ & ${(R\ts{n}^*)^2}/{D^*\ts{n,typ}}$ & Diffusion timescale in the negative electrode solid material & $2.564 \times 10^3$ \\
    $\tau^*\ts{p}$ & ${(R^*\ts{p})^2}/{D^*\ts{p,typ}}$ & Diffusion timescale in the positive electrode solid material & $1 \times 10^3$ \\
    $\tau^*\ts{e}$ & ${(L^*_x)^2}/{D^*\ts{e,typ}}$ & Diffusion timescale in the electrolyte & $1.816 \times 10^2$ \\
    $\tau_{r\ts{n}}^*$ & ${F^*}/({m^*\ts{n,typ} a^*\ts{n} (c^*\ts{e,typ})^{1/2}})$ & Reaction timescale in the negative electrode & $8.475 \times 10^2$\\
    $\tau_{r\ts{p}}^*$ & ${F^*}/({m^*\ts{p,typ} a^*\ts{p} (c^*\ts{e,typ})^{1/2}})$ & Reaction timescale in the positive electrode & $3.390 \times 10^4$\\
    $\tau^*\ts{th}$ & ${\rho^*\ts{eff} (L^*)^2}/{\lambda^*\ts{eff}}$ & Planar ($y$--$z$) thermal diffusion timescale & $5.724 \times 10^2$ \\
    \bottomrule
    \end{tabular}}
        \caption{Timescales associated with the physical processes occurring in the battery model. We note that the discharge timescale is not quite the same as the discharge time (otherwise its value would be $3600/{\mathcal C}$, since a 1C discharge takes 3600 seconds). The difference arises from using $c^*\ts{n,max} L^*_x (L^*)^2$ as the scale for the available lithium, which differs from the actual capacity in four respects: (i) the thickness of the negative electrode is $L^*_n$ not $L^*_x$; (ii) the cross-sectional area of the current collectors is not  $(L^*)^2$; (iii) the volume of  active material is lower than the superficial volume, by a factor that depends on the surface area per unit volume, the size of the active particle, and the porosity; (iv) the theoretical capacity is lower than the actual capacity. The OCP tends to $\pm \infty$ when the battery is fully charged/discharged; the actual  capacity is that available between two (finite) voltage limits, which we take to be \SI{3.2}{V} and \SI{4.7}{V} in defining ${\mathcal C}$}
            \label{table:timescales}
\end{table}


\section{Comparison of the COMSOL and PyBaMM Numerical Solutions in 1D}
\label{subsection:1D}
Here we give a brief comparison of the numerical solution of the standard 1D DFN model obtained using PyBaMM with that from COMSOL, to give an estimate of the effect of  the different in spatial discretisation methods and time-stepping routines used by the two software packages. 

\begin{figure}
    \centering
    \includegraphics[width=\textwidth]{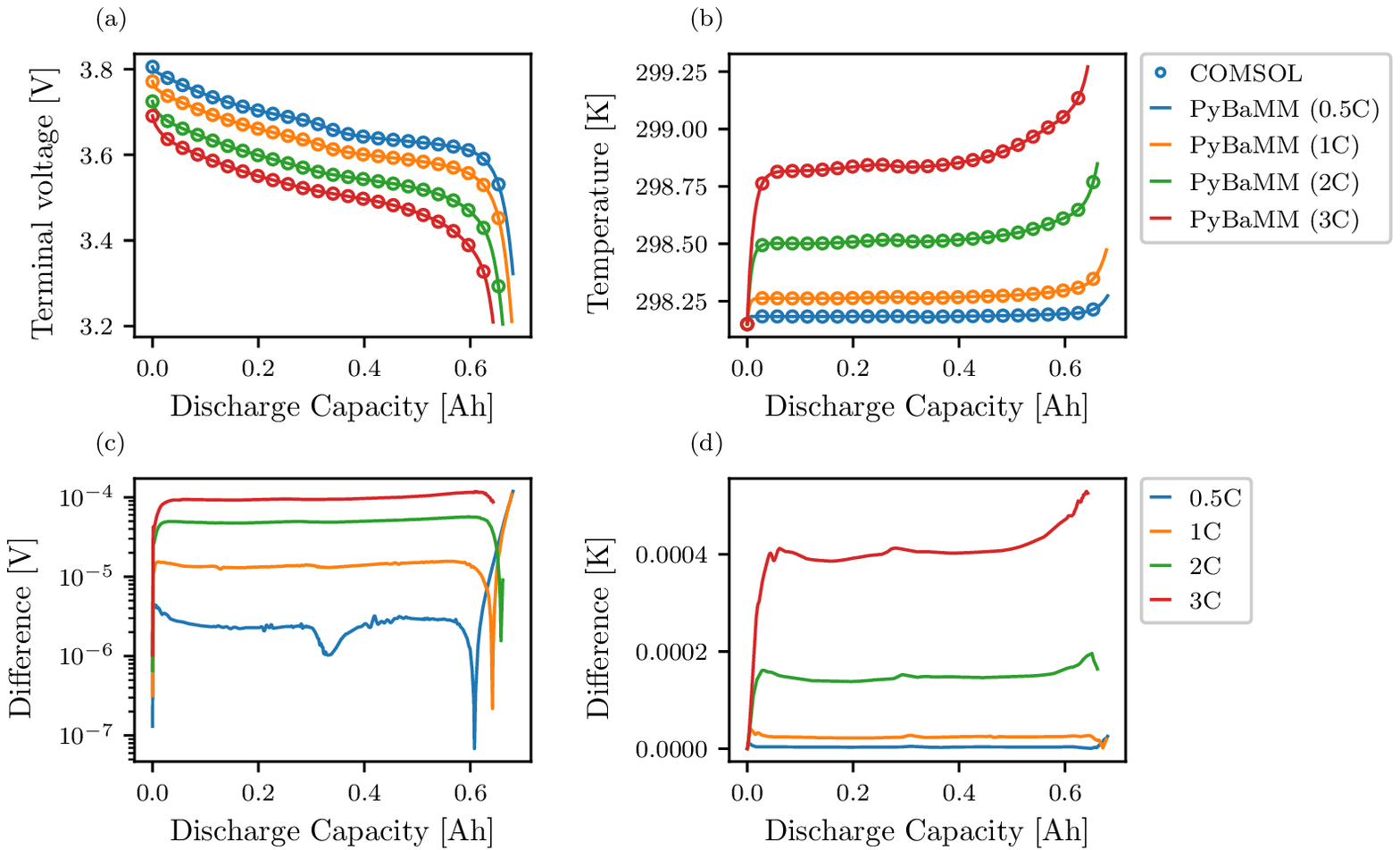}
    \caption{Comparison of (a) the terminal voltage and (b) the temperature obtained from the solution of the 1D DFN model from PyBaMM and COMSOL for a range of C-rates. The differences in the voltage and temperature between the two solutions are shown in panels (c) and (d), respectively. Note the logarithmic scale in panel (c). Here we used 128 finite volumes per domain in the PyBaMM model and the COMSOL model was solved on an ``extremely fine'' mesh (45 elements in each electrode, 11 elements in the separator). Both time stepping routines used a relative tolerance of $10^{-6}$.}
    \label{fig:1D_voltage_temperature_comparison}
\end{figure}

\begin{figure}[p]
    \centering
    \includegraphics[width=\textwidth]{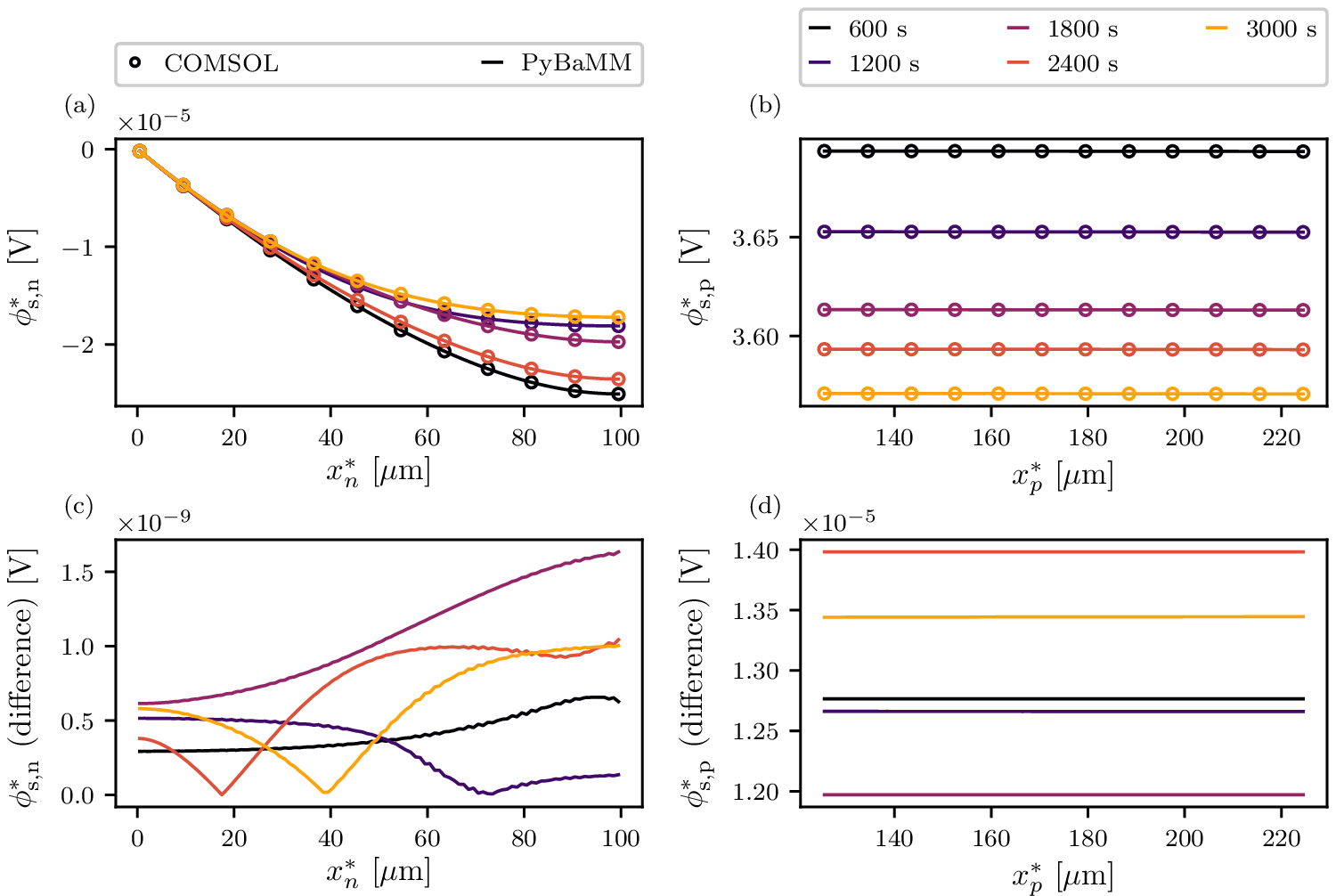}
    \caption{Comparison of the solid potential in the (a) negative electrode and (b) positive electrode obtained from the solution of the 1D DFN model from PyBaMM and COMSOL. The differences in the calculated potentials in the negative and positive electrodes are shown in panels (c) and (d), respectively.}
    \label{fig:1D_phi_s_comparison}

    \centering
    \includegraphics[width=\textwidth]{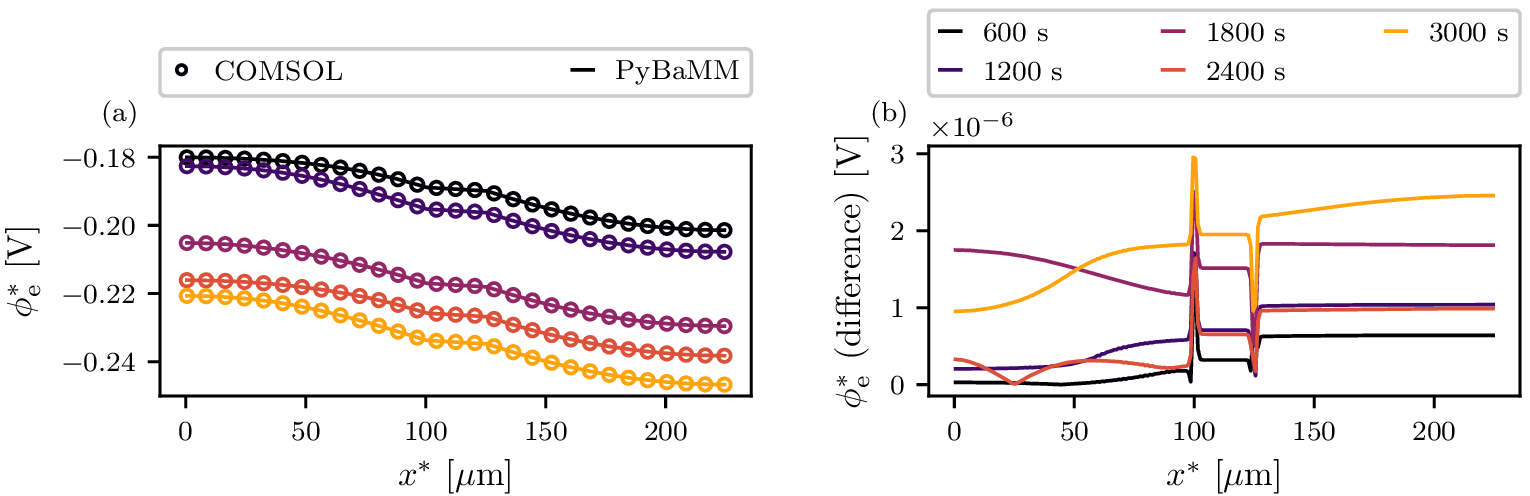}
    \caption{(a) Comparison of the electrolyte potential obtained from the solution of the 1D DFN model from PyBaMM and COMSOL. (b) The difference in the calculated potential.}
    \label{fig:1D_phi_e_comparison}
\end{figure}

\begin{figure}[p]
    \centering
   \includegraphics[width=\textwidth]{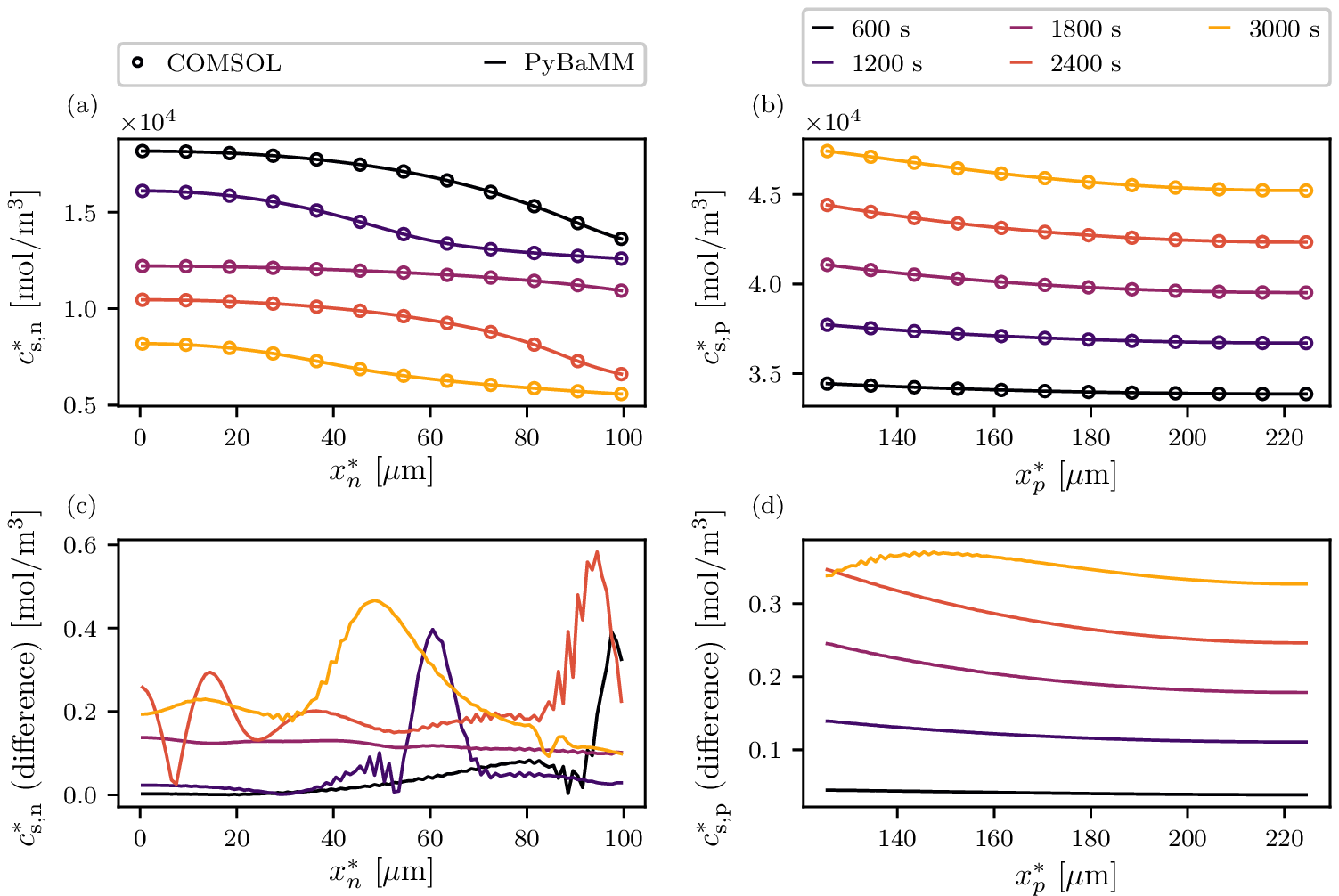}
    \caption{Comparison of the surface concentrations in the (a) negative electrode particles and (b) positive electrode particles obtained from the solution of the 1D DFN model from PyBaMM and COMSOL. The differences in the calculated concentrations in the negative and positive electrodes are shown in panels (c) and (d), respectively.}
    \label{fig:1D_c_s_comparison}

    \centering
   \includegraphics[width=\textwidth]{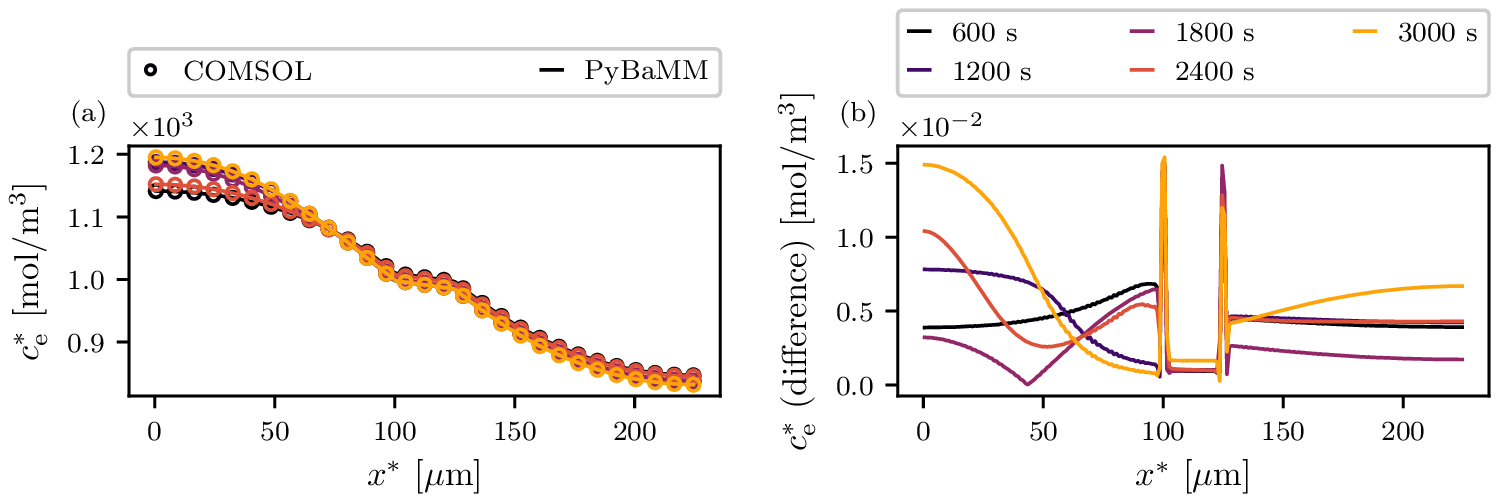}
    \caption{(a) Comparison of the electrolyte concentration obtained from the solution of the 1D DFN model from PyBaMM and COMSOL. (b) The difference in the calculated concentration.}
    \label{fig:1D_c_e_comparison}
\end{figure}

In Figure~\ref{fig:1D_voltage_temperature_comparison}, we present the computed terminal voltage and volume-averaged temperature across a range of C-rates. Panels (c) and (d) show the differences between the two solutions. It can be seen that the difference in the terminal voltage is no more than $\mathcal{O}(10^{-4})$ \si{V}. Similarly, the difference in the computed average cell temperature is also at most $\mathcal{O}(10^{-4})$ \si{K.}. At low to moderate C-rates, a large spike in the difference is observed towards the end of the simulation where the OCV in highly non-linear. Both meshes were refined until the solutions converged, but the differences still persisted, which leads us to believe the discrepancy is due to slight differences in the time-stepping strategy and approach used to couple the micro- and macro-scale problems. Despite this, the errors are sufficiently small to provide us with confidence that any errors larger than this are a result of the asymptotic reduction and not differences in the solution approach. Further comparisons of the potentials and concentrations through the cell thickness are provided in Figures~\ref{fig:1D_phi_s_comparison}--\ref{fig:1D_c_e_comparison}.

\begin{table}[htb]
	\centering
    \resizebox{\textwidth}{!}{
	\begin{tabular}{c c c c c c c c c c}
	\toprule
     $N$ & $\phi\ts{s,n}^*$ & $\phi\ts{s,p}^* - V^*$& $\phi\ts{e}^*$  & $c\ts{s,n,surf}^*$  & $c\ts{s,p,surf}^*$ & $c\ts{e}^*$ & $V^*$ & $\tilde{T}^*$ &  Solution time [s] \\
     \midrule
        4 &\num{2.56e-2}  & \num{6.59e-2} & \num{6.06e-3} & \num{8.62e-3} & \num{3.72e-3} & \num{2.53e-3} & \num{4.60e-3} & \num{2.86e-5} & 0.1712\\
        8 &\num{6.86e-3}  & \num{1.64e-2} & \num{1.75e-3} & \num{2.57e-3} & \num{9.51e-4} & \num{6.47e-4} & \num{1.37e-3} & \num{8.53e-5} & 0.1925\\
        16 &\num{2.05e-3} & \num{4.11e-3} & \num{6.32e-4} & \num{9.43e-4} & \num{2.56e-4} & \num{1.75e-4} & \num{4.04e-4} & \num{2.45e-6} &  0.3027\\
        32 &\num{8.78e-4} & \num{1.03e-3} & \num{3.48e-4} & \num{5.50e-4} & \num{8.26e-5} & \num{6.30e-5} & \num{1.40e-4} & \num{7.74e-7} & 0.6790\\
        64 &\num{6.27e-4} & \num{2.71e-4}& \num{2.78e-4} & \num{4.64e-4} & \num{4.00e-5} & \num{4.26e-5} & \num{7.43e-5} & \num{3.60e-7} & 2.119\\
        128 &\num{5.76e-4}& \num{1.08e-4} & \num{2.61e-4} & \num{4.45e-4} & \num{2.98e-5} & \num{3.87e-5} & \num{6.80e-5} & \num{2.59e-7} & 7.49\\
    \bottomrule
	\end{tabular}
	} 
    \caption{Normalised RMS difference between the PyBaMM and COMSOL solution of the 1D DFN model for a selection of model variables. The tabulated quantities for a variable $\psi$ were computed as RMS($\psi\ts{PyBaMM}-\psi\ts{COMSOL}$)/RMS($\psi\ts{COMSOL}$). Here $N$ is the number of finite volumes per domain in the PyBaMM model. The PyBaMM solution was compared to the COMSOL solution on an ``extremely fine" mesh (45 elements in each electrode, 11 elements in the separator). Both time stepping routines used a relative tolerance of $10^{-6}$.}
    \label{table:1D_errors_vs_npts}
\end{table}

To investigate the influence of the mesh size in the PyBaMM implementation of the DFN  model a number of simulations were performed with an increasing number of finite volumes per domain. Table~\ref{table:1D_errors_vs_npts} shows the normalised root mean square (RMS) error in the potentials, concentrations, voltage and temperature obtained by solving the model in PyBaMM as the number of finite volumes $N$ is increased. The RMS error was computed with respect to the solution obtained using COMSOL's ``extremely fine" mesh (45 elements in each electrode, 11 elements in the separator), which was typically solved in around 70s.

\section{Summary of reduced models in dimensional form}
Here we re-dimensionalise the reduced models we have derived, dropping the superscripts $^{(0)}$ for ease of notation.
\subsection{The high conductivity limit}
\label{sec:high_dim}
Using $\Omega^*$ to denote the cross-section of the cell we have, in the current collectors,
\begin{subequations}
  \label{eqn:high_summary_dim}
\begin{equation}
    L\ts{cn}^* \sigma^*\ts{cn} \nablap^2\phi\ts{s,cn}^* = \mathcal{I}^*,  
    \qquad
    L\ts{cp}^*\sigma^*\ts{cp}\nablap^2 \phi\ts{s,cp}^* = -\mathcal{I}^* \qquad \mbox{ in }  \Omega^*
\end{equation}
with boundary conditions
\begin{equation}
  \phi\ts{s,cn}^* = 0 \quad \mbox{ on } \partial \Omega^*\ts{tab,cn,$\perp$}, \qquad
\nablap \phi\ts{s,cn}^*\cdot\boldsymbol{n} = 0 \quad \mbox{ on } \partial \Omega^*\ts{ext,cn,$\perp$}  
\end{equation}
\begin{equation}
  -\sigma^*\ts{cp}\nablap \phi^*\ts{s,cp}  \cdot\boldsymbol{n}
  = \frac{I^*_{\text{app}} }{  A^*\ts{tab,cp}}  \  \mbox{ on } \partial\Omega\ts{tab,cp,$\perp$},\qquad
\nablap \phi\ts{s,cp}^*\cdot\boldsymbol{n} = 0 \  \mbox{ on } \partial\Omega^*\ts{ext,cp,$\perp$},
\end{equation}
where $\mathcal{I}^*$ is the through-cell current given (at each point $(y^*,z^*)\in \Omega^*$) by a one-dimensional DFN model (or any alternative one-dimensional through-cell battery model which gives the current $\mathcal{I}^*$ in terms of the current collector potentials $ \phi\ts{s,cn}^*$ and $\phi\ts{s,cp}^*$ and temperature $T^*$).

This electrical problem is coupled to the two-dimensional thermal problem
\begin{align}
\qquad  \rho\ts{eff}^*  \pdv{T^*}{t^*} &=  \lambda\ts{eff}^* \nablap^2 T^* +  \bar{Q}^* - \frac{(h^*\ts{cn}+h^*\ts{cp}) }{(L^*\ts{cn} + L\ts{cp}^*+L_x^*) }    (T^* -T^*_\infty)\quad \mbox{ in }\Omega,\\
    - \lambda\ts{eff}^*\nablap T^* \cdot \boldsymbol{n} &=  h^*\ts{eff}    (T^* -T^*_\infty) \quad \mbox{ on }\partial\Omega,
\end{align}
where the heat source is
\begin{align}
  \bar{Q}^* & = \frac{1}{(L_x^*+L^*\ts{cn}+L^*\ts{cp})}\left(\int_0^{L_x^*}Q^*\ts{DFN}\, \dd{x^*}  + L^*\ts{cn} \sigma^*\ts{cn}|\nablap \phi\ts{s,cn}^*|^2
  +
  L^*\ts{cp}  \sigma^*\ts{cp}  |\nablap \phi\ts{s,cp}^*|^2\right),
\end{align}
where $Q^*\ts{DFN} = Q^*\ts{Ohm,k} + Q^*\ts{rxn,k} + Q^*\ts{rev,k}$ ($\kin{n,s,p}$) is the heat source in the through-cell  one-dimensional DFN model, the effective edge heat transfer coefficient is
\[
h^*\ts{eff} = \frac{1}{(L_x^*+L^*\ts{cn}+L^*\ts{cp})} \int_{-L^*\ts{cn}}^{L_x^*+L^*\ts{cp}} h^*   \, \dd{x^*},
\]
and the effective volumetric heat capacity  and thermal conductivity are
\begin{equation}
    \rho^*\ts{eff} = \frac{\sum\limits\ts{k} \rho^*\ts{k} c^*\ts{p,k} L^*\ts{k}}{\sum\limits\ts{k} L^*\ts{k}}, \qquad \lambda^*\ts{eff} = \frac{\sum\limits\ts{k} \lambda^*\ts{k} L^*\ts{k}}{\sum\limits\ts{k} L^*\ts{k}},
\end{equation}
respectively.

\end{subequations}

\subsubsection{The very high conductivity limit}
\label{sec:very_high_dim}
\begin{subequations}
In the very high conductivity limit the model reduces further, to 
    \label{model_very_high_dim}
    \begin{align}
  \label{eq:very_high_V_dim}
    V^* & =  V^*\ts{DFN}({\mathcal I}^*, T^*)- 
  R^*\ts{cp}I^*\ts{app}
  -
  R^*\ts{cn} I^*\ts{app},\\
  \label{eq:very_high_T_dim}
  \rho^*\ts{eff} \pdv{T^*}{t^*} &=
\bar{Q}^*\ts{DFN}({\mathcal{I}}^*, T^*)
- \frac{(h^*\ts{cn}+h^*\ts{cp})  }{ (L^*\ts{cn}+L^*\ts{cp}+L_x^*)}  (T^*-T^*_\infty)\\ \nonumber
& \mbox{ }-
\frac{(T^*-T^*_\infty)}{L^*_y L^*_z} \int_{\partial \Omega^*}h^*\ts{eff}\, \dd{s^*} +H^*\ts{cn} (I^*\ts{app})^2
+H^*\ts{cp} (I^*\ts{app})^2 \qquad \mbox{ in }\Omega^*,
    \end{align}
    where the current density 
    \[{\mathcal I}^* = \frac{I^*\ts{app}}{|\Omega^*|},\]
    where $|\Omega^*|$ is the area of $\Omega^*$,
and  the effective current collector resistances are given by
\begin{equation}  R^*\ts{cn}= 
  \frac{ \av{f^*\ts{n}}}{L^*_y L^*_z L^*\ts{cn}\sigma^*\ts{cn}},
  \qquad  R^*\ts{cp} =
  \frac{  1}{L^*_y L^*_z 
 A^*\ts{tab,cp}\sigma^*\ts{cp}}  \int_{\partial\Omega^*\ts{tab,cp,$\perp$}}   f^*\ts{p}\, \dd{s^*},
\end{equation}
and the power coefficients $H^*\ts{cn}$ and $H^*\ts{cp}$ are given by
    \begin{equation}
  H^*\ts{cn} =\frac{L^*\ts{cn}\av{|\nablap f^*\ts{n}|^2}}{(L_x^*+L^*\ts{cn}+L^*\ts{cp})(L^*_y L^*_z L^*\ts{cn})^2\sigma^*\ts{cn}} ,\quad
     H^*\ts{cp} =
     \frac{L^*\ts{cp}   \av{|\nablap f^*\ts{p}|^2}}{(L_x^*+L^*\ts{cn}+L^*\ts{cp})(L^*_y L^*_z L^*\ts{cp})^2 \sigma^*\ts{cp}},
\end{equation}
where $f^*\ts{n}$ and $f^*\ts{p}$ satisfy the auxiliary problems
\begin{equation}
   \nablap^2 f^*\ts{n}=  -1, \qquad  \qquad 
   \nablap^2 f^*\ts{p} = 1 \qquad \mbox{ in }  \Omega^*,
\end{equation}
\begin{align}
  f^*\ts{n} &= 0 \quad \mbox{ on } \partial \Omega^*\ts{tab,cn,$\perp$},& 
\nablap f^*\ts{n}\cdot\boldsymbol{n} &= 0 \quad \mbox{ on } \partial \Omega^*\ts{ext,cn,$\perp$}  ,\\
\nablap f^*\ts{p}\cdot\boldsymbol{n} &=
\frac{L^*_y L^*_zL^*\ts{cp}}{ A^*\ts{tab,cp}}  \  \mbox{ on } \partial\Omega^*\ts{tab,cp,$\perp$},&
\nablap f^*\ts{p}\cdot\boldsymbol{n} &= 0 \
\mbox{ on } \partial\Omega^*\ts{ext,cp,$\perp$}, \qquad \av{f^*\ts{p}}=0.
\end{align}
These steady two-dimensional problems depend only on the geometry of the current collectors and the position of the tabs, and can be solved
independently of (\ref{eq:very_high_V_dim}) and (\ref{eq:very_high_T_dim}).
The term
\begin{align}
  \bar{Q}^*\ts{DFN}({\mathcal{I}}^*, T^*) & = \frac{1}{(L_x^*+L^*\ts{cn}+L^*\ts{cp})}\int_0^{L_x^*}Q^*\ts{DFN}\, \dd{x^*} \end{align}
is the heat generated  in the electrodes, electrolyte and separator from the one-\linebreak dimensional through-cell DFN problem.

After solving this single one-dimensional model,  the potential distribution in the current collectors may be evaluated as 
\begin{equation}
  \label{eq:very_high_phi_cc_dim}
 \phi^*\ts{s,cn} = -\frac{ \mathcal{I}^*}{ L^*\ts{cn}\sigma^*\ts{cn}} f^*\ts{n} , \qquad
\phi^*\ts{s,cp} = V^* + \frac{ \mathcal{I}^*}{ L^*\ts{cp}\sigma^*\ts{cp}} f^*\ts{p}  .
\end{equation}

\end{subequations}

\clearpage
\bibliographystyle{plain}
\bibliography{refs}